\documentclass[12pt]{iopart}
\usepackage{graphicx}
\usepackage{iopams}
\usepackage{xcolor,soul}

\begin{document}

\title[]{Relation between topology and heat currents in multilevel absorption machines}

\author{J Onam Gonz{\'a}lez, Jos{\'e} P Palao and Daniel Alonso}

\address{Departamento de F{\'i}sica and IUdEA, Universidad de La Laguna, La Laguna 38204, Spain}

\eads{dalonso@ull.es}

\begin{abstract}

The steady state heat currents of continuous absorption machines can be decomposed into thermodynamically consistent contributions, each of them associated with a circuit in the graph representing the master equation of the thermal device. We employ this tool to study the functioning of absorption refrigerators and heat transformers with an increasing number of active levels. Interestingly, such an analysis is independent of the particular physical implementation (classical or quantum) of the device. We provide new insights into the understanding of scaling up thermal devices concerning both the performance and the magnitude of the heat currents. Indeed, it is shown that the performance of a multilevel machine is smaller or equal than the corresponding to the largest circuit contribution. Besides, the magnitude of the heat currents is well-described by a purely topological parameter which in general increases with the connectivity of the graph. Therefore, we conclude that for a fixed number of levels, the best of all different constructions of absorption machines is the one whose associated graph is as connected as possible, with the condition that the performance of all the contributing circuits is equal.
\end{abstract}

\pacs{1315, 9440T}



\maketitle

\section{Introduction}


Continuous quantum absorption machines \cite{Kosloff2014} are multilevel systems connected to several thermal baths at different temperatures. Their autonomous functioning can be rigorously described by using the theory of open quantum systems \cite{Breuer2002}. Some basic models such as the three-level \cite{Palao2001,Linden2010}, the two-qubit \cite{Levy2012} and the three-qubit \cite{Linden2010,Correa2013} absorption refrigerators have been widely employed in establishing fundamental relations in quantum thermodynamics \cite{Kosloff2013}. Besides, several experimental proposals have been put forward, for example those based on nano-mechanical oscillators or atoms interacting with optical resonators \cite{Mari2012,Mitchison2016}, atoms interacting with nonequilibrium electromagnetic fields \cite{Leggio2015}, superconducting quantum interference devices \cite{hofer2016,Chen2012}, and quantum dots \cite{Venturelli2013}. Further, an experimental realization of a quantum absorption refrigerator has been recently reported \cite{Maslennikov2017}.

The dynamics of quantum machines is described by a master equation when the coupling with the baths is weak enough. Along this paper we consider in addition systems for which two states with the same energy cannot be connected to a third one through the same bath. This assumption greatly simplifies the quantum master equation as the population and coherence dynamics are decoupled in the system energy eigenbasis \cite{Breuer2002}, and  will be referred in the following as the PCD condition. It guarantees the thermodynamic consistency of the models \cite{Spohn1977,Alicki1979}, which may be broken when some uncontrolled approximations are introduced \cite{Levy2014}. Under this assumption coherences decay with time and are irrelevant in the steady state functioning of the device, contrary, for example, to externally driven devices \cite{Uzdin2015} and systems including matter currents \cite{Cuetara2016}, where they may play an important role on the thermodynamic properties. When the PCD condition holds, the populations follow a continuous time Markov master equation \cite{Gardiner1985}, given in terms of the rates for the transitions between states, which are always allowed in both directions. In this case a thermal device implemented in a quantum system can be described within the framework of stochastic thermodynamics \cite{Seifert2012,Broek2015,Esposito2009}.

The analysis of the thermodynamic quantities can be realized at different levels of description \cite{Polettini2016}. From a macroscopic point of view, where the relevant quantities are the bath temperatures and the physical (total) heat currents, to a microscopic description that considers in addition the device structure. This latter perspective is more and more relevant as the advance of the experimental techniques allows for the design and the manipulation of the device. A prominent tool for this microscopic analysis is graph theory, where the stochastic master equation for the populations is represented by a graph. Schnakenberg theory \cite{Schnakenberg1976} is a popular approach that gives a decomposition of the total entropy production based on a set of fundamental circuits in the graph. Basically, Schnakenberg applies Kirchhoff's current laws to reduce the number of terms appearing in the entropy production, which may be highly beneficial for optimization procedures. It has been used for example in linear irreversible thermodynamics \cite{Yamamoto2016} and in the study of steady-state fluctuation theorems \cite{Andrieux2007,Rahav2013}. This method does not intend to associate an entropy production with each circuit. In particular, the attempt to interpret individually each term in the decomposition may lead to apparent negative entropy productions, although this problem can be avoided by a convenient choice of the fundamental circuits \cite{MacQueen1981,Kalpazidou2007,Jiang2004}. However, it has been shown that the diagnosis of the machine performance greatly benefits from considering the thermodynamic analysis of not only the fundamental but all the possible circuits in the graph \cite{Einax2014,Correa2015,Gonzalez2016}. A convenient approach is then Hill theory \cite{Hill1966}. Schnakenberg and Hill theory assign the same affinity to each circuit, but the latter considers all the possible circuits and leads to thermodynamically consistent entropy productions. Both methods coincide when the fundamental set of circuits contains all the possible ones.

In this paper we use Hill theory to fully characterize the two relevant quantities in the study of continuous absorption devices: the steady state heat currents and performance. Our aim is to find out under what conditions these quantities are as large as possible, i.e. what is the best construction of multilevel machines. Graph theory allows us to answer this key question from a very general perspective, looking only at the topological structure of the graph. Although we are motivated by the study of quantum models and in the following we will assume the PCD condition, our analysis also applies to classical stochastic models, including mesoscopic systems where the relevant degrees of freedom are identified by a coarse graining procedure \cite{Esposito2012,Altaner2012a}. In fact, the main advantage of this approach is that many properties of a device can be inferred from its graph representation irrespectively of its underlying, microscopic or mesoscopic, quantum or classical, realization.

It has been shown that systems with degenerate energy levels and driven by an external field may present a linear increment of the heat currents with the number of states \cite{Gelbwaser2015,Niedenzu2015}. Furthermore, two-stroke models in the quasi-equilibrium regime show an improvement in the performance with the number of levels \cite{Silva2016}. However, using a particular construction of continuous absorption devices by merging three-level systems, Correa \cite{Correa2014c} found no changes in the performance and a fast saturation in the magnitude of the heat currents as the number of levels increases. Thus the question arises whether this limitation may be overcome by different designs of the absorption device.

We are interested in continuous machines that either extract energy from the coldest bath (refrigerators) or inject energy to the hottest bath (heat transformers). We do not consider devices designed for complicated tasks involving more than one target bath, although our procedure could also be applied to such systems. The best refrigerators and heat transformers should generally provide the largest possible heat currents and be also able of reaching the reversible limit for a particular set of the parameters. In order to identify them, we first justify that a machine coupled to three baths is capable of achieving the same currents than more complicated devices which consider additional heat reservoirs. As multilevel machines are composed by multiple circuits, our next step is to identify the optimal circuit to be used as building block. In general the magnitude of the heat currents increases with the transitions rates for any circuit. Hence, to elucidate the role of the circuit structure in the currents we set the rates to fixed values. Moreover, this condition avoids processes which prevents the machine from reaching the reversible limit when considering multiple circuits. The following step is to determine the graph structure leading to the largest heat currents considering optimal blocks. Finally, we relax the condition of fixed rates to improve the scaling of the currents with the number of levels without introducing harmful processes as heat leaks.

The paper is organized as follows: in section \ref{sec:motivation} we motivate the generic nature of our work by describing two different models of absorption devices which are represented by the same four-state graph. The master equation for all the quantum models used as illustration of the general results can be obtained using \ref{sec:appendixrates} with the Hamiltonians provided in \ref{sec:appendixHamiltonian}. We also introduce in section \ref{sec:motivation} the essential concepts of graph theory needed to characterize the heat currents associated with a circuit inside a general graph. This result allows us to relate each circuit to a thermodynamic mechanism and classify it attending to its contribution to the overall functioning of a device coupled to three baths. Although we have used previously the circuit decomposition in a different context, the analysis of the irreversible mechanisms arising in  thermal devices indirectly connected to environments \cite{Gonzalez2016}, we provide now a derivation of it using Hill theory in \ref{sec:appendixdecomposition}. The differences between Hill and Schnakenberg decompositions are discussed and worked out for the four-state graph in \ref{sec:appendixfourstates} and \ref{sec:appendixSchnakenberg}. In section \ref{sec:circuit} we analyze multilevel machines represented by a graph circuit. Explicit expressions for the scaling of the heat currents with the number of levels in the high and low temperature limits are provided in \ref{sec:appendixR}. Machines represented by graphs with multiple circuits are studied in section \ref{sec:multiplecircuits}. A simple example to illustrate the relation between the heat currents and the graph connectivity is presented in \ref{sec:appendixG3B}. We draw our conclusions in section \ref{sec:conclusion}.

\section{Motivation and background}\label{sec:motivation}

\begin{figure}[h]
\centering
\begin{minipage}{\linewidth}\centering
\includegraphics[width=\linewidth]{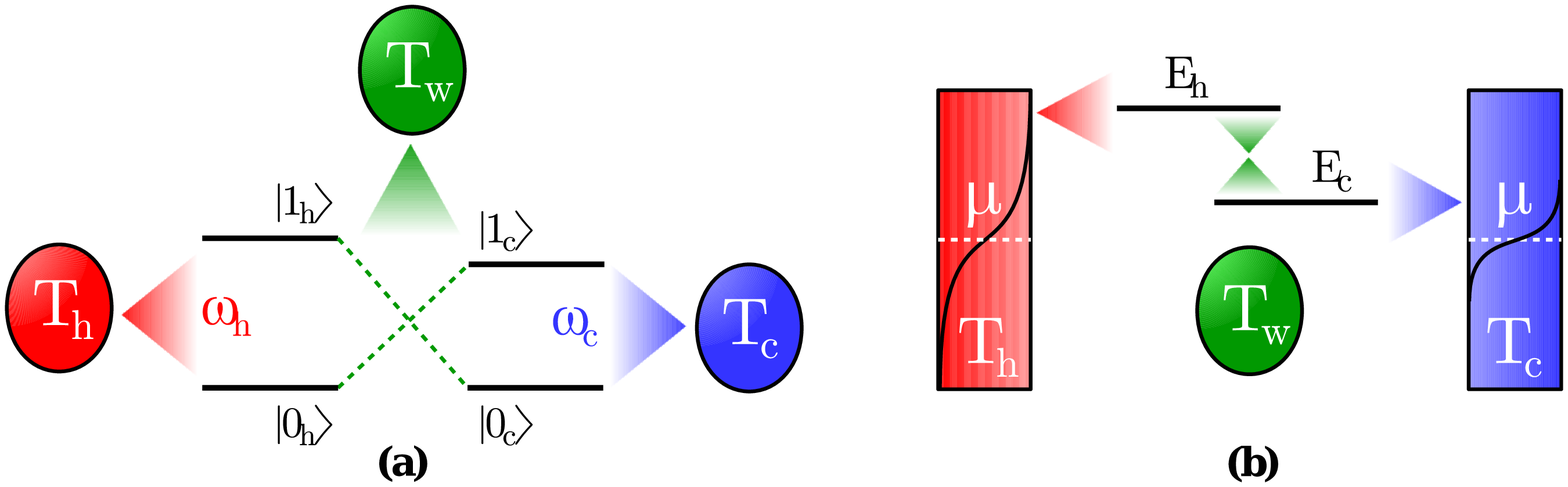}
\end{minipage}

\vspace*{0.5cm}

\includegraphics[width=0.89\linewidth]{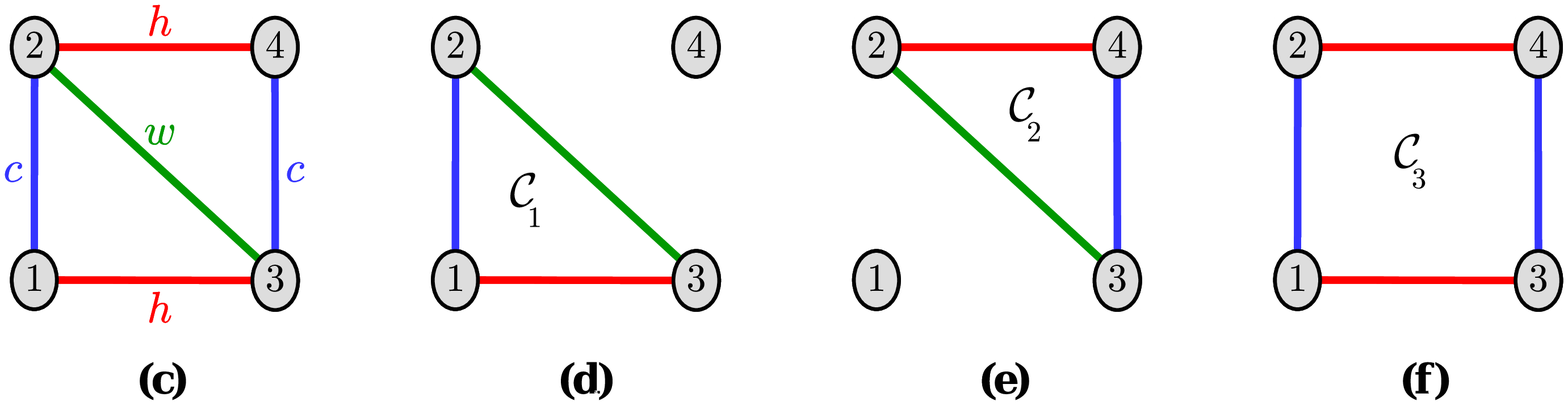}
\caption{Schematic representation of (a) a two-qubit device and (b) a photoelectric device. (c) The same graph, $\mathcal{G}_4$, represents the master equation associated with each one. Three circuits can be indentified: (d) $\mathcal{C}_1$, (e) $\mathcal{C}_2$ and (f) $\mathcal{C}_3$.}\label{fig:fig1}
\end{figure}

We will motivate our approach by first considering two different models of absorption devices, both connected to three reservoirs at temperatures $T_c$ (cold), $T_h$ (hot) and $T_w$ (referred in the following as the temperature of the work bath, in analogy with devices driven by an external field), with $T_c<T_h<T_w$. Depending on internal parameters, the devices can either work as heat transformers, transferring energy from the hot to the work bath, or as refrigerators, extracting energy from the cold bath assisted by the work bath. The first model is the two-qubit device \cite{Levy2012} shown in figure \ref{fig:fig1}(a). Each qubit is connected to a bosonic heat bath at temperatures $T_c$ and $T_h$. The interaction between them is mediated by another bath at temperature $T_w$. The state of this machine can be expanded in the product state basis $|1\rangle\equiv|0_h0_c\rangle$, $|2\rangle\equiv|0_h1_c\rangle$, $|3\rangle\equiv|1_h0_c\rangle$ and $|4\rangle\equiv|1_h1_c\rangle$, with energies $E_1=0$, $E_2=\hbar\omega_c$, $E_3=\hbar\omega_h$ and $E_4=\hbar(\omega_c+\omega_h)$. When the system is weakly coupled to the reservoirs, the dynamics of the populations is described by a master equation 
\begin{equation}\label{eq:master4}
\frac{d}{dt} p_i(t) = \sum_{j=1}^{4}(W_{ij}^c+W_{ij}^w+W_{ij}^h)\, p_j(t)\,,
\end{equation}
where $p_i$ are the populations, $W_{ij}^\alpha\ge 0$ the transition rates associated with the coupling with the bath $\alpha$, and $W^\alpha_{ii} = -\sum_{j\neq i}W^\alpha_{ji}$. For our purpose now is only important that the non-zero rates associated with the cold bath correspond to transitions $1\leftrightarrow 2$, $3\leftrightarrow 4$, with the hot bath to $1\leftrightarrow 3$, $2\leftrightarrow 4$, and with the work bath to $2\leftrightarrow 3$.

The second model is a photoelectric device \cite{Einax2014,Esposito2009b,Cleuren2012,Li2013} composed of two single-level, spinless quantum dots with energies $E_c$ and $E_h$ that can be both occupied at the same time. Each dot is connected to a metal electrode with  chemical potential $\mu<E_c<E_h$ and temperatures $T_c$ and $T_h$,  see figure \ref{fig:fig1}(b). We choose the same chemical potential in order to avoid introducing mechanical work and assume that neither the temperatures nor the chemical potential are modified by the interchange of electrons through the quantum dots. The system states are $1\equiv 0_h0_c$, $2\equiv 0_h1_c$, $3\equiv 1_h0_c$ and $4\equiv 1_h1_c$, with energies $E_1=0$, $E_2=E_c$, $E_3=E_h$ and $E_4=E_c+E_h$, where now $0_\alpha$ and $1_\alpha$ are the number of electrons in the dot $\alpha$. Transitions between the two dots are supported by an additional radiation source (for example the Sun in photovoltaic models \cite{Einax2014,Li2013}) at an effective temperature $T_w$. Considering a weak coupling with the electrodes and a negligible line broadening of the energy levels, the device dynamics can be described by an stochastic master equation \cite{Averin1991} in the form (\ref{eq:master4}), but with transition rates determined by the particularities of the physical model under consideration. In the case of the absorption device with bosonic baths the rates are proportional to Planck distributions, while for the photoelectric device they are proportional to Fermi functions.

The relevant point for our analysis is that since the master equations have the same structure, the devices share several thermodynamic properties that stem directly from it. The different physics involved in each case is only reflected in the particular values of the transition rates. The master equation may be represented by a network, a weighted and labeled multi-digraph. However, as transition between states are always allowed in both directions, we will use a simpler representation consisting in a labeled graph, with vertices associated with the system states and undirected edges with the transitions  \cite{Schnakenberg1976,Hill1966}. When necessary, an arbitrary orientation can be assigned to the graph and a weight to each edge, given by the corresponding transition rate. For example, the representation of (\ref{eq:master4}), denoted in the following by $\mathcal{G}_4$, is shown in figure \ref{fig:fig1}(c). The circuits of $\mathcal{G}_4$, defined as a cyclic sequence of distinct edges, are displayed in figure \ref{fig:fig1}(d), (e) and (f). Circuits $\mathcal{C}_1$ and $\mathcal{C}_2$ participate in different processes depending on their two possible orientations, referred as cycles. For example the cycle $\vec{\mathcal{C}}_1\equiv \{1,2,3,1\}$ absorbs energy from the cold and work baths that is rejected to the hot bath, whereas the opposite cycle $-\vec{\mathcal{C}}_1\equiv \{1,3,2,1\}$ absorbs energy from the hot bath and rejects it to the cold and work baths. In both processes there is a net exchange of energy with the three baths. 

The circuit $\mathcal{C}_3$ involves only two baths (cold and hot) and in our models does not lead to any net exchange of energy. This is a consequence of having the same energy gap for transitions assisted by the same bath. However, in more general setups with different transition energies, $E_{34}=E_{12}+\Delta$ and $E_{24}=E_{13}+\Delta$ with $E_{ji}=E_i-E_j$, there is a heat leak which increases with the energy shift $\Delta$ \cite{Correa2015}.

The overall physical heat currents $\dot{\mathcal{Q}}_\alpha$ and the performance of the device are then the result of the interplay of the different mechanisms related to each circuit. In spite of the simplicity of the previous qualitative interpretation, the microscopic currents $\dot{q}_\alpha(\mathcal{C}_\nu)$ corresponding to each circuit in the graph are not straightforwardly obtained from the physical currents. We introduce below the concepts of graph theory needed to characterize them.


\subsection{Graph, circuits and steady state heat currents}

For simplicity we consider systems with $N$ states of energies $E_i$, $1 \le i \le N$, represented by a connected graph and coupled with thermal baths. The generalization for systems exchanging particles without involving any mechanical work, as the absorption device of figure \ref{fig:fig1}(b), is straightforward. The system transitions may be coupled to one or several independent heat baths, each one in equilibrium at temperature $T_\alpha$, $1 \le \alpha \le R$.  The system evolution is described by a master equation 
\begin{equation}\label{eq:master}
\frac{d}{dt} p_i(t) = \sum_{j=1}^{N}\sum_{\alpha=1}^R W_{ij}^\alpha\, p_j(t)\,,
\end{equation}
where $p_i$ is the normalized probability distribution to be in the state $i$, $W_{ij}^\alpha\ge 0$ is the transition rate from the state $j$ to the state $i$ due to the coupling with the bath $\alpha$, and
\begin{equation}\label{eq:diag}
W^\alpha_{ii} = -\sum_{j\neq i}W^\alpha_{ji}\,.
\end{equation}
The transition matrix $\mathbf{W}$, with elements $W_{ij}=\sum_{\alpha=1}^R\, W_{ij}^\alpha$, is singular, which guarantees the existence of a non-trivial steady state solution of (\ref{eq:master}) and the conservation of the normalization. In addition, we assume that
\begin{equation}\label{eq:nondiag}
\frac{W_{ji}^\alpha}{W_{ij}^\alpha} = \exp\left[\frac{E_{ji}}{k_B T_\alpha}\right]\,,
\end{equation}
where $k_B$ is the Boltzmann constant.  If the transition rates $W_{ij}^\alpha$ for $j>i$ are known, the remaining rates can be determined by using  (\ref{eq:diag}) and (\ref{eq:nondiag}).

\begin{figure}[h]
\centering
\includegraphics[width=\linewidth]{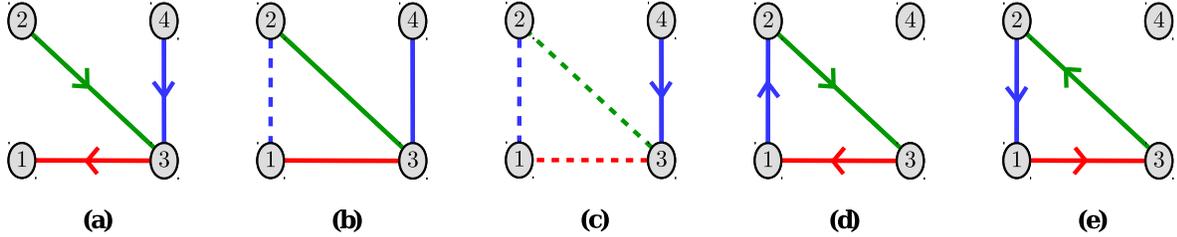}
\caption{(a) A maximal tree $\mathcal{T}^1$ of $\mathcal{G}_4$  oriented towards the vertex 1 is denoted by $\vec{\mathcal{T}}^1_1$. (b) When adding the chord $x_1$ (dashed line) to $\mathcal{T}^1$, the circuit $\mathcal{C}_1$ is obtained. (c) Removing the circuit (dashed line) and orienting the remaining edges towards it, the forest $\vec{\mathcal{F}}_1^1$ is found. The cycles $\vec{\mathcal{C}}_1$ and $-\vec{\mathcal{C}}_1$ are shown in (d) and (e).}
\label{fig:fig2}
\end{figure}

The master equation (\ref{eq:master}) is represented by a graph $\mathcal{G}(N,U)$ composed of $N$ vertices and $U$ undirected edges.
Let $x_e$ be an edge in the graph, $1 \le e \le U$. In the following $\vec{x}_e$ will denote an edge oriented from vertex $i_e$ to $j_e$, whereas $-\vec{x}_e$ connects $j_e$ to $i_e$, in both cases due to the coupling with the bath $\alpha_e$. Oriented edges are related to rate coefficients by $W(\vec{x}_e)=W_{j_ei_e}^{\alpha_e}$ and $W(-\vec{x}_e)=W_{i_ej_e}^{\alpha_e}$. An algebraic value $\mathcal{A}$ may be assigned to any oriented subgraph $\vec{\mathcal{G}}_s$ of $\mathcal{G}$, composed of $s\le U$ oriented edges $\vec{x}_e$\cite{Schnakenberg1976},
\begin{equation}\label{eq:algebraic1}
\mathcal{A}(\vec{\mathcal{G}}_s) = \prod_{\alpha=1}^R\,\mathcal{A}^\alpha(\vec{\mathcal{G}}_s)\,,
\end{equation}
where, if the subgraph involves edges associated with the bath $\alpha$,
\begin{equation}\label{eq:algebraic2}
\mathcal{A}^\alpha(\vec{\mathcal{G}}_s) = \prod_{e \in s,\alpha}\,W(\vec{x}_e)\,,
\end{equation}
with $\prod_{e \in s,\alpha}$the product over all the directed edges of $\vec{\mathcal{G}}_s$ corresponding to this bath, and otherwise $\mathcal{A}^\alpha(\vec{\mathcal{G}}_s)=1$. Both $\mathcal{A}(\vec{\mathcal{G}}_s)$ and $\mathcal{A}^\alpha(\vec{\mathcal{G}}_s)$ are positive real numbers. A maximal tree $\mathcal{T}^\mu$ , $1 \le \mu \le N_T$, is a subgraph of $\mathcal{G}$ containing $N-1$ edges without forming any closed path. The oriented subgraph $\vec{\mathcal{T}}_i^{\mu}$ is a maximal tree in which all the edges are directed towards the vertex $i$. A chord of a maximal tree is one of the $U-N+1$ edges that are not part of it. The subgraph obtained when a chord is added to a maximal tree has only a circuit $\mathcal{C}_\nu$, $1 \le \nu \le N_C$. When removing the circuit from the previous subgraph, a collection of edges remains. Orienting them towards the circuit, a forest $\vec{\mathcal{F}}_\nu^\beta$ is found. The index $\beta$ indicates that for a given circuit different forests can be found, resulting from different maximal tress. The number of maximal trees ($N_T$), circuits ($N_C$) and forests depend on the topological structure of the graph $\mathcal{G}$. Each circuit $\mathcal{C}_\nu$ may be oriented in one of the two possible directions, leading to the cycles $\vec{\mathcal{C}}_\nu$ and $-\vec{\mathcal{C}}_\nu$. Some examples are shown in figure \ref{fig:fig2}. In \ref{sec:appendixdecomposition} we use Hill theory to show that the steady state heat current associated with a circuit is given by
\begin{equation}\label{eq:cycleheatcurrent}
\dot{{q}}_\alpha(\mathcal{C}_\nu)
= -T_\alpha D(\mathcal{G})^{-1}\det(-\mathbf{W}|\mathcal{C}_\nu)
[\mathcal{A}(\vec{\mathcal{C}}_\nu) - \mathcal{A}(-\vec{\mathcal{C}}_\nu)]X^\alpha(\vec{\mathcal{C}}_\nu)
 \,.
\end{equation}
The factor $D$ is calculated using
\begin{equation}\label{eq:determinant}
D(\mathcal{G})=\sum_{i=1}^N\sum_{\mu=1}^{N_T}\,\mathcal{A}(\vec{\mathcal{T}}_i^{\mu})=
|\det(\widetilde{\mathbf{W}})|\,.
\end{equation}
The quantity $D(\mathcal{G})>0$  increases with the complexity (both the number of vertices and edges). It is a factor which reduces the population in a circuit and therefore the corresponding heat currents when considering machines with an increasing number of them. The matrix $\widetilde{\mathbf{W}}$ is obtained from the transition matrix ${\mathbf{W}}$ by replacing the elements of an arbitrary row by ones, whereas the matrix $(-\mathbf{W}|\mathcal{C}_\nu)$ is obtained by removing from $-\mathbf{W}$ all the rows and columns corresponding to the vertices of the circuit. Indeed, $\det(-\mathbf{W}|\mathcal{C}_\nu)$ is the sum of the forests of $\mathcal{C}_\nu$ and can be thought of as an ``injection of population'' through edges not belonging to it.

We have also introduced the cycle affinity associated with the bath $\alpha$,
\begin{equation}\label{eq:cycleaffinitybath}
X^\alpha(\vec{\mathcal{C}}_\nu) = k_B \ln\left(\frac{\mathcal{A}^\alpha(\vec{\mathcal{C}}_\nu)}
{\mathcal{A}^\alpha(-\vec{\mathcal{C}}_\nu)}\right)\,,
\end{equation}
and then the total cycle affinity is 
\begin{equation}\label{eq:cycleaffinity}
X(\vec{\mathcal{C}}_\nu) = \sum_{\alpha=1}^R X^\alpha(\vec{\mathcal{C}}_\nu) = 
k_B \ln\left(\frac{\mathcal{A}(\vec{\mathcal{C}}_\nu)}
{\mathcal{A}(-\vec{\mathcal{C}}_\nu)}\right)\,.
\end{equation}
The quantity $-T_\alpha X^\alpha(\vec{\mathcal{C}}_\nu)$ is just the net amount of energy interchanged between the bath $\alpha$ and the system when performing the cycle $\vec{\mathcal{C}}_\nu$. Notice that $X^\alpha(-\vec{\mathcal{C}}_\nu)=-X^\alpha(\vec{\mathcal{C}}_\nu)$ and hence each cycle is related to a process where some energy is either absorbed from or rejected to the bath. The circuit heat current (\ref{eq:cycleheatcurrent}) can be viewed as the result of the competition between the two cycles, described by $-T_\alpha [\mathcal{A}(\vec{\mathcal{C}}_\nu) - \mathcal{A}(-\vec{\mathcal{C}}_\nu)]X^\alpha(\vec{\mathcal{C}}_\nu)$, weighted by how the circuit is immersed in the graph, which is contained in $D(\mathcal{G})^{-1}\det(-\mathbf{W}|\mathcal{C}_\nu)$.

As a consequence of (\ref{eq:nondiag})
\begin{equation}\label{eq:Xcondition}
\sum_{\alpha=1}^R T_\alpha X^\alpha(\vec{\mathcal{C}}_\nu) = 0\,,
\end{equation}
reflecting that the net energy exchanged by the system with the baths along a complete cycle is zero. Using it the following relation is found,
\begin{equation}\label{eq:heatcurrentcondition}
\sum_{\alpha=1}^R\dot{q}_\alpha(\mathcal{C}_\nu) = 0\,,
\end{equation}
and since the only contribution to the steady state entropy production is due to finite-rate heat transfer effects, the circuit entropy production is
\begin{equation}\label{eq:cycleentropy2}
\dot{s}(\mathcal{C}_\nu) = -\sum_{\alpha=1}^R
\frac{\dot{q}_\alpha(\mathcal{C}_\nu)}{T_\alpha}\geq 0\,,
\end{equation}
where the inequality is shown in \ref{sec:appendixdecomposition}. These two last equations assure the consistency of the circuit heat currents and the entropy production with the first and second laws of thermodynamics. Finally, the total entropy production is given by $\dot{S} = \sum_{\nu=1}^{N_C}\dot{s}(\mathcal{C}_\nu)$ and the physical heat currents by $\dot{\mathcal{Q}}_\alpha = \sum_{\nu=1}^{N_C}\dot{q}_\alpha(\mathcal{C}_\nu)$. They can be directly obtained from the transition rates by using (\ref{eq:cycleheatcurrent}), without determining the steady state populations. As an illustration of the circuit decomposition, the heat currents for the graph $\mathcal{G}_4$ are  worked out in \ref{sec:appendixfourstates}. Let us remind that other  decompositions of $\dot{S}$ are possible and we briefly discuss them in \ref{sec:appendixSchnakenberg}.

The circuit heat currents (\ref{eq:cycleheatcurrent}) are homogeneous functions of degree 1 with respect to the transition rates, that is
\begin{equation}\label{eq:scaling}
W_{ij}^\alpha \rightarrow \sigma W_{ij}^\alpha\,;\hspace*{1cm}
\dot{q}_\alpha(\mathcal{C}_\nu) \rightarrow \sigma \dot{q}_\alpha(\mathcal{C}_\nu)\,;\hspace*{1cm}
\dot{\mathcal{Q}}_\alpha \rightarrow \sigma \dot{\mathcal{Q}}_\alpha\,,
\end{equation}
with $\sigma>0$. Therefore, the currents can be always modified by changing the rates, provided that the assumptions to obtain the master equation remain valid. This property emphasizes the importance of the graph topology.

\subsection{Classification of circuits}

The contribution of each circuit $\mathcal{C}_\nu$ to the physical heat currents can be classified attending to their non-zero affinities $X^\alpha$:

\begin{itemize}

\item[(i)] $X^\alpha(\vec{\mathcal{C}}_\nu)=0$ for all the baths. These circuits will be referred as trivial circuits, as they do not contribute neither to the steady state heat currents nor to the entropy production.

\item[(ii)] Condition (\ref{eq:Xcondition}) prevents any circuit from having only a non-zero affinity $X^\alpha$.

\item[(iii)]  $X^\alpha(\vec{\mathcal{C}}_\nu)\neq 0$ only for two baths, $\alpha=\alpha_1,\alpha_2$. Then there is only a net energy transfer between them, although other baths could participate in the cycle. Using (\ref{eq:heatcurrentcondition}) and (\ref{eq:cycleentropy2}), the following condition is found
\begin{equation}
\dot{q}_{\alpha_1}(\mathcal{C}_\nu)\left(\frac{1}{T_{\alpha_2}}-\frac{1}{T_{\alpha_1}}\right)\ge 0\,.
\end{equation}
Taking $T_{\alpha_1}<T_{\alpha_2}$, the heat currents verify $\dot{q}_{\alpha_2}(\mathcal{C}_\nu)>0$ and $\dot{q}_{\alpha_1}(\mathcal{C}_\nu)<0$. Therefore the net heat current associated with these circuits always flows from the higher temperature bath to the lower temperature one. In the context of refrigerators and heat transformers these circuits are related to heat leaks that decrease the performance \cite{Correa2015,Gonzalez2016}. 

\item[(iv)] $X^\alpha(\vec{\mathcal{C}}_\nu)\neq 0$ for three baths, $\alpha=\alpha_1,\alpha_2,\alpha_3$. They will be referred as three-bath circuits in the following. Equation (\ref{eq:Xcondition}) implies that, given a circuit orientation, two of the affinities and their corresponding heat currents must have the same sign. Considering ${\rm sgn}(X^{\alpha_1})={\rm sgn}(X^{\alpha_2})=-{\rm sgn}(X^{\alpha_3})$ and using again (\ref{eq:heatcurrentcondition}) and (\ref{eq:cycleentropy2}), we obtain
\begin{equation}
\dot{q}_{\alpha_1}(\mathcal{C}_\nu)\left(\frac{1}{T_{\alpha_3}}-\frac{1}{T_{\alpha_1}}\right)
+\dot{q}_{\alpha_2}(\mathcal{C}_\nu)\left(\frac{1}{T_{\alpha_3}}-\frac{1}{T_{\alpha_2}}\right)
\ge 0\,.
\end{equation}
\end{itemize}

The formalism applies also to circuits with non-zero affinities associated with more than three baths, but they are not relevant for our analysis.

\subsection{Circuits in refrigerators and heat transformers}\label{sec:subcircuits}

For simplicity we discuss now refrigerators, but the results are also valid for heat transformers. In general, the environment may be composed by the target coldest bath, a collection of sink baths with temperatures $\{T_{h,i}\}$ (where the surplus energy is rejected) and work baths with temperatures $\{T_{w,i}\}$ (supplying energy to complete the cycles). Let $\{X^{\alpha,i}\}$ be the affinities of a particular circuit. Equation (\ref{eq:cycleheatcurrent}) implies that we can always find a hot and a work bath with temperatures and affinities given by $T_\alpha X^\alpha\equiv \sum_i T_{\alpha,i}X^{\alpha,i}$ ($\alpha=h,w$), such that tuning their rate values (\ref{eq:scaling}) we obtain the same or larger heat currents than in the original system. Therefore we focus in the following on circuits and thermal machines coupled to three thermal baths with temperatures $T_c<T_h<T_w$.

In the construction of the device we do not consider circuits with two edges associated with different baths connecting the same vertices, as it would lead directly to heat leaks (iii). To perform useful tasks we must include three-bath circuits (iv), which can be classified as:

\begin{itemize} 
\item[(a)] $\alpha_1=h$ and $\alpha_2=w$, which leads to $\dot{q}_{h}(\mathcal{C}_\nu),\dot{q}_{w}(\mathcal{C}_\nu)>0$ and $\dot{q}_{c}(\mathcal{C}_\nu)<0$. 

\item[(b)] $\alpha_1=c$ and $\alpha_2=h$, giving now $\dot{q}_{c}(\mathcal{C}_\nu),\dot{q}_{h}(\mathcal{C}_\nu)<0$ and $\dot{q}_{w}(\mathcal{C}_\nu)>0$. 

\item[(c)] $\alpha_1=c$ and $\alpha_2=w$, for which 
\begin{equation}\label{eq:conditionaffinity}
{\rm sgn}[X^{c}(\vec{\mathcal{C}}_\nu)] = {\rm sgn}[X^{w}(\vec{\mathcal{C}}_\nu)]=-{\rm sgn}[X^{h}(\vec{\mathcal{C}}_\nu)]\,.
\end{equation}

\end{itemize}

In cases (a) and (b) heat is simply transferred from the work to the cold bath, whereas the hot bath absorbs or gives up some energy. In (c) two different directions for the heat currents are possible: $\dot{q}_{c}(\mathcal{C}_\nu),\dot{q}_{w}(\mathcal{C}_\nu)<0$, $\dot{q}_{h}(\mathcal{C}_\nu)>0$  and $\dot{q}_{c}(\mathcal{C}_\nu),\dot{q}_{w}(\mathcal{C}_\nu)>0$, $\dot{q}_{h}(\mathcal{C}_\nu)<0$, which correspond to the conditions for the heat currents in heat transformers and refrigerators respectively. Therefore equation (\ref{eq:conditionaffinity}) settles the condition for the affinities in useful circuits. The particular working mode will depend on the system parameters.

\section{Thermal machines represented by a circuit graph}\label{sec:circuit}

In this section we analyze thermal machines that are represented by a circuit graph, $\mathcal{G}=\mathcal{C}^{N}$, with $N\ge 3$ states (vertices) and $U=N$ undirected edges.  We consider useful three-bath circuits for which (\ref{eq:conditionaffinity}) holds. Along this section we shall make explicit the circuit length (the number of states or edges) by the superscript $N$. In this case the physical and circuit heat currents coincide. From (\ref{eq:cycleaffinity}) we obtain $\mathcal{A}(-\vec{\mathcal{C}}^{\,N})=\mathcal{A}(\vec{\mathcal{C}}^{\,N})\exp[-X(\vec{\mathcal{C}}^{\,N})/k_B]$, and then the physical heat currents are given by
\begin{equation}\label{eq:onecircuitheatcurrent}
\dot{\mathcal{Q}}_\alpha = \dot{q}_\alpha(\mathcal{C}^N)
= -T_\alpha
\mathcal{R}(\vec{\mathcal{C}}^{\,N})
\{1-\exp[-X(\vec{\mathcal{C}}^{\,N})/k_B]\}
X^\alpha(\vec{\mathcal{C}}^{\,N})\,,
\end{equation}
where
\begin{equation}\label{eq:R}
\mathcal{R}(\vec{\mathcal{C}}^{\,N})=D(\mathcal{C}^N)^{-1}\mathcal{A}(\vec{\mathcal{C}}^{\,N})\,.
\end{equation}
Notice that the dependence on the arrangement of the edges in the circuit is contained in $\mathcal{R}(\vec{\mathcal{C}}^{\,N})$ and the currents vanish for $X(\vec{\mathcal{C}}^{\,N})=0$. Using (\ref{eq:Xcondition}), the circuit affinity is rewritten as
\begin{equation}
X(\vec{\mathcal{C}}^{\,N}) =
\left(1-\frac{T_c}{T_w}\right) X^c(\vec{\mathcal{C}}^{\,N})+
\left(1-\frac{T_h}{T_w}\right) X^h(\vec{\mathcal{C}}^{\,N})\,.
\end{equation}

The device operating mode depends only on the parameter $x= -(T_cX^c)/(T_hX^h)$, $0\leq x\leq 1$, which is independent of the particular circuit orientation, and for $X(\vec{\mathcal{C}}^{\,N})=0$ results in
\begin{equation}
x_r = \frac{T_c(T_w-T_h)}{T_h(T_w-T_c)}\,.
\end{equation}
When $x<x_r$, the device operates as an absorption refrigerator whose coefficient of performance is
\begin{equation}\label{eq:COP}
\varepsilon = \frac{\dot{\mathcal{Q}}_c}{\dot{\mathcal{Q}}_w} = \frac{x}{1-x}\,.
\end{equation}
The coefficient of performance reaches the Carnot value $\varepsilon_C=T_c(T_w-T_h)/[T_w(T_h-T_c)]$ when $x$ approaches to $x_r$ from below but at vanishing heat currents ($X(\vec{\mathcal{C}}^{\,N})=0$). When $x>x_r$, the machine operates as a heat transformer with efficiency 
\begin{equation}\label{eq:efficiency}
\eta = \frac{-\dot{\mathcal{Q}}_w}{\dot{\mathcal{Q}}_h}
= 1-x\,,
\end{equation}
reaching the Carnot value $\eta_C=T_w(T_h-T_c)/[T_h(T_w-T_c)]$ when $x$ approaches to $x_r$ from above. In consequence, the device performance depends only on the circuit affinities $X^\alpha(\vec{\mathcal{C}}^{\,N})$, irrespective of the value $\mathcal{R}(\vec{\mathcal{C}}^{\,N})$, and they may be suitably tuned to reach the reversible limit for any graph circuit.

\subsection{Circuit structure, performance and heat currents}

In the following and without loss of generality, we choose a circuit orientation such that $X(\vec{\mathcal{C}}^{\,N})>0$. The affinities and the algebraic value $\mathcal{A}(\vec{\mathcal{C}}^{\,N})$ depend only on the number of edges and their associated transitions rates. In particular, $\mathcal{A}(\vec{\mathcal{C}}^{\,N})$ is the product of $N$ transition rates. However, the factor $D$ depends also on the arrangement of the edges through the oriented maximal trees in (\ref{eq:determinant}). The $N_T=N$ maximal trees are obtained by removing in each case one of the edges in the circuit. We denote by $\mathcal{T}^j$ the maximal tree obtained by removing the edge starting in the state $j$. The term $D(\mathcal{C}^N)$ is the sum of $N^2$ terms $\mathcal{A}(\vec{\mathcal{T}}_i^j)$, each one composed of the product of $N-1$ transition rates. 

\subsubsection{Dependence on the transition rates}

\begin{figure}[h]
\centering
\includegraphics[width=0.45\linewidth]{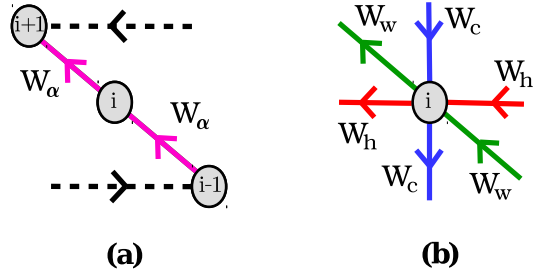}
\caption{(a) The state $i$ in a circuit is connected to $i-1$ and $i+1$ by the same bath. The PCD condition ($E_{i-1}\neq E_{i+1}$) requires  $E_{i-1}<E_{i}<E_{i+1}$ or $E_{i-1}>E_{i}>E_{i+1}$. Then, for a given circuit orientation, a path from $i-1$ to $i+1$ consists in two jumps either absorbing energy from or rejecting energy to the bath. In both cases the transition rate $W_\alpha$ is the same. (b) When considering more general graphs, the PCD condition implies that the maximum number of edges connecting a state is six in a machine connected to three baths.}
\label{fig:fig3}
\end{figure}

From the previous results for $\mathcal{A}$ and $D$ and after a straightforward calculation, the heat currents are bounded by
\begin{equation}\label{eq:heatcurrentbound}
|\dot{\mathcal{Q}}_\alpha|<T_\alpha W_m 
|X^\alpha(\vec{\mathcal{C}}^{\,N})|\,,
\end{equation}
where $W_m$ is the minimum rate in $\mathcal{A}(\vec{\mathcal{C}}^{\,N})$. As intuitively expected, increasing the lowest rates may result in larger heat currents for any circuit. The remaining question is then what kind of circuit shows the largest heat currents for a set of fixed transition rates. In order to answer it, we assume in the following that the available resource in the machine design is a set of three undirected edges with fixed transitions rates, $W_\alpha$ and $W_{-\alpha}$, associated  the first with energy transfer to and the second with energy absorption from the bath $\alpha=c,w,h$. This construction can always overcome complicated ones with more that three edges using a proper scaling of the rates, see (\ref{eq:scaling}). Besides, it implies fixed energy gaps $|E_{ij}|\equiv E_{\alpha}$ for transitions assisted by the same bath and:

\begin{itemize}

\item[(i)] When two edges, $x_{i-1}$ and $x_i$, connecting the state $i$ are associated with the same bath, then $W(\vec{x}_{i-1})=W(\vec{x}_i)$ for any of the two cycles as a consequence of the PCD condition, see figure \ref{fig:fig3}(a). 

\item[(ii)] The minimum number of edges required to construct a useful three-bath circuit is three, therefore $E_c+E_w=E_h$ as a result of (\ref{eq:Xcondition}) and  (\ref{eq:conditionaffinity}). For simplicity we take $E_c\neq E_w$. 

\end{itemize}

\begin{figure}[h]
\centering
\begin{minipage}{0.32\linewidth}
\includegraphics[width=\linewidth]{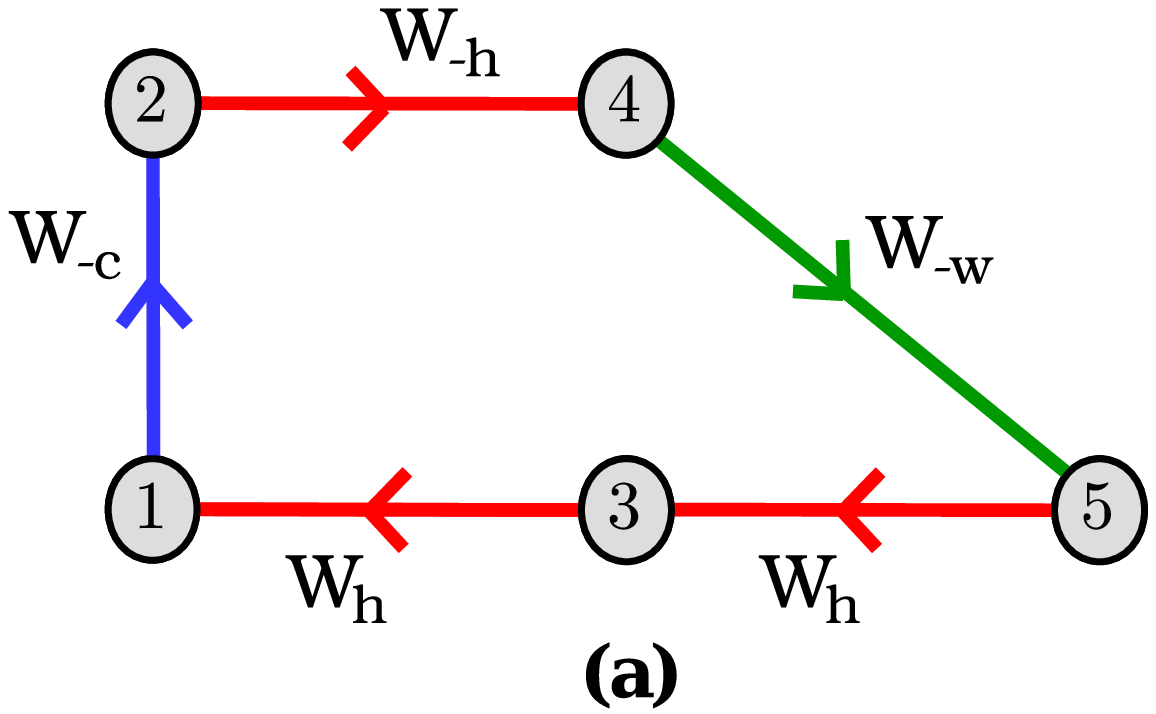}
\end{minipage}\hfill
\begin{minipage}{0.32\linewidth}
\includegraphics[width=\linewidth]{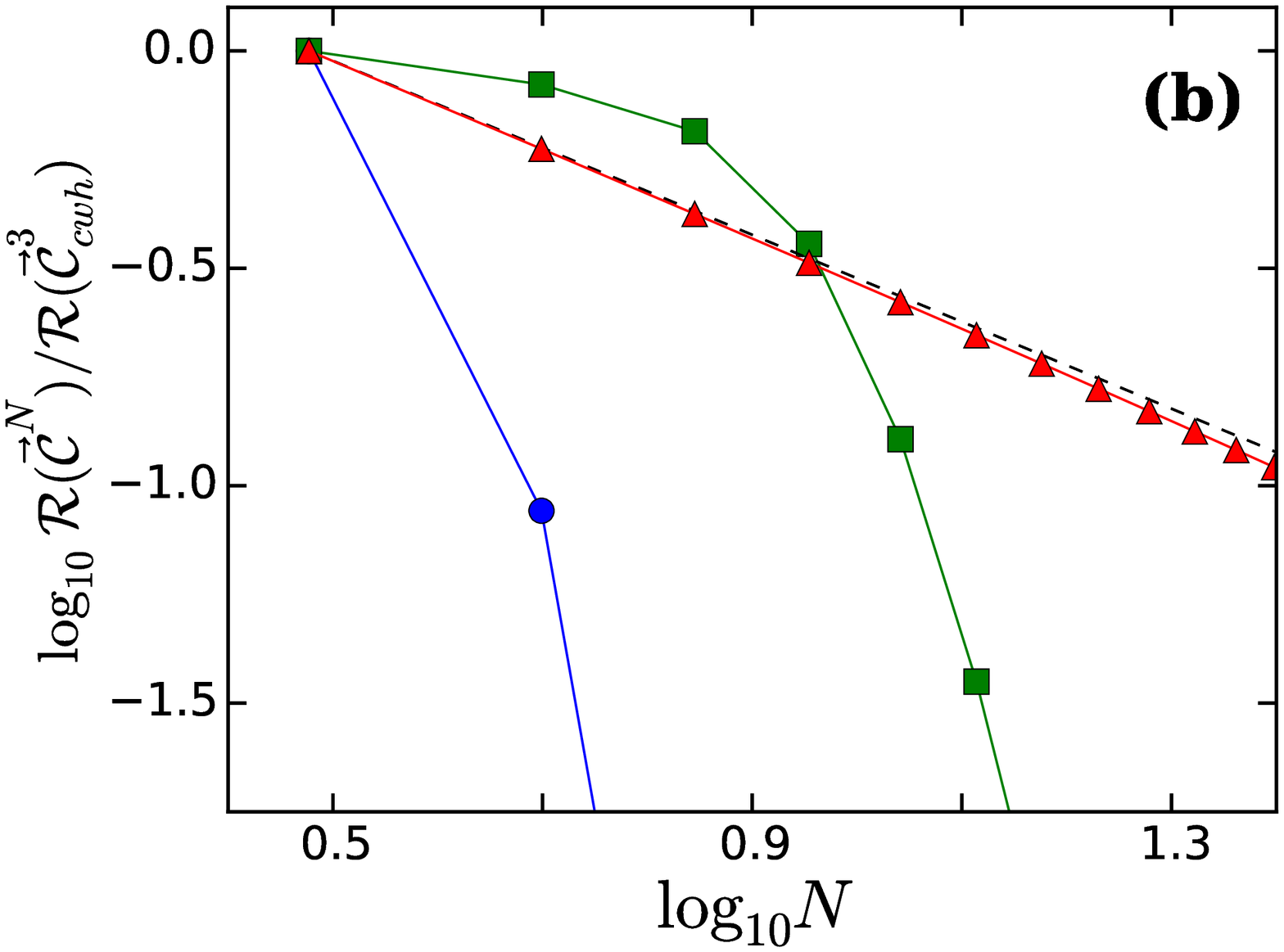}
\end{minipage}\hfill
\begin{minipage}{0.32\linewidth}
\includegraphics[width=\linewidth]{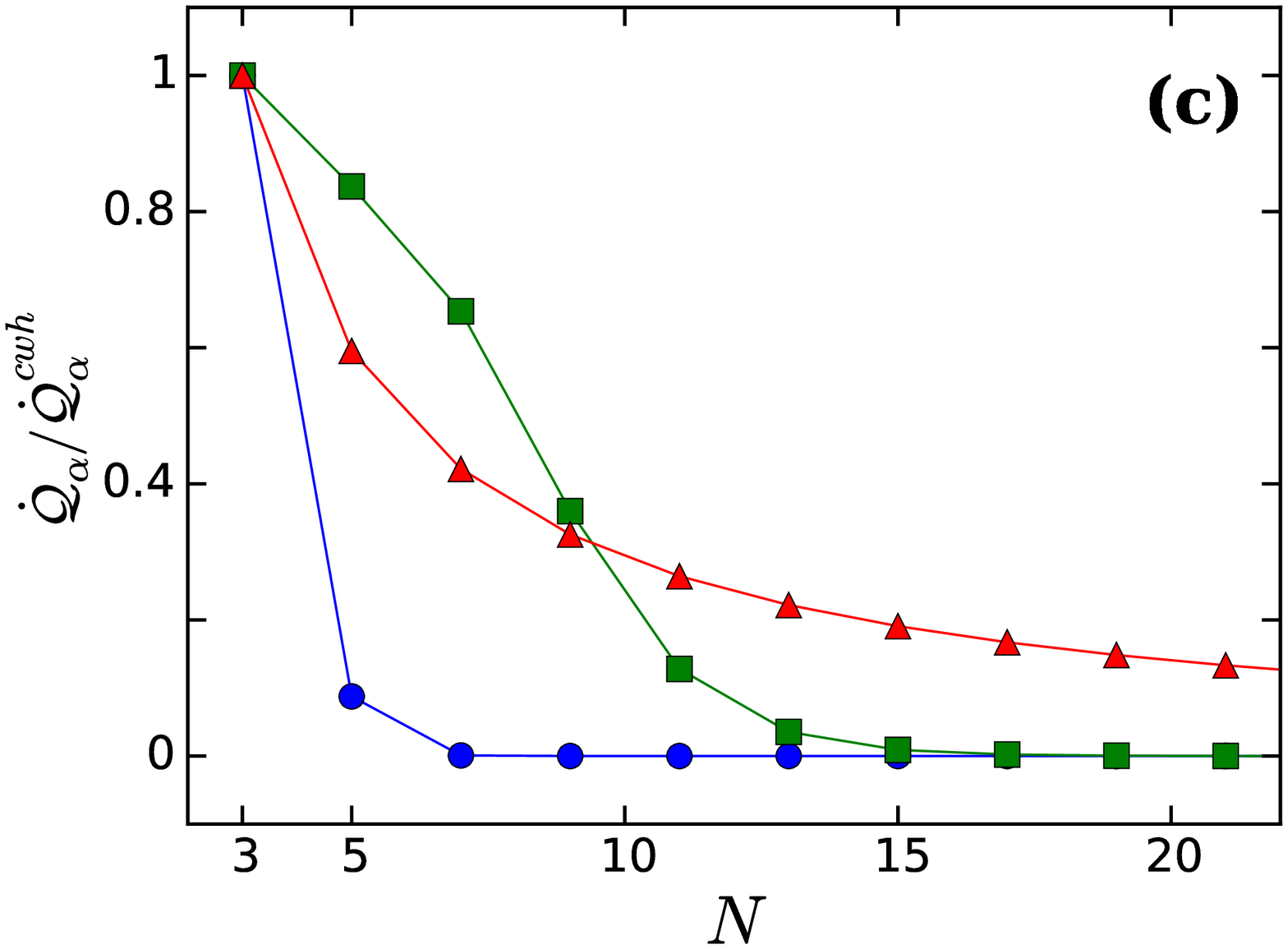}
\end{minipage}
\begin{minipage}{0.32\linewidth}
\includegraphics[width=0.9\linewidth]{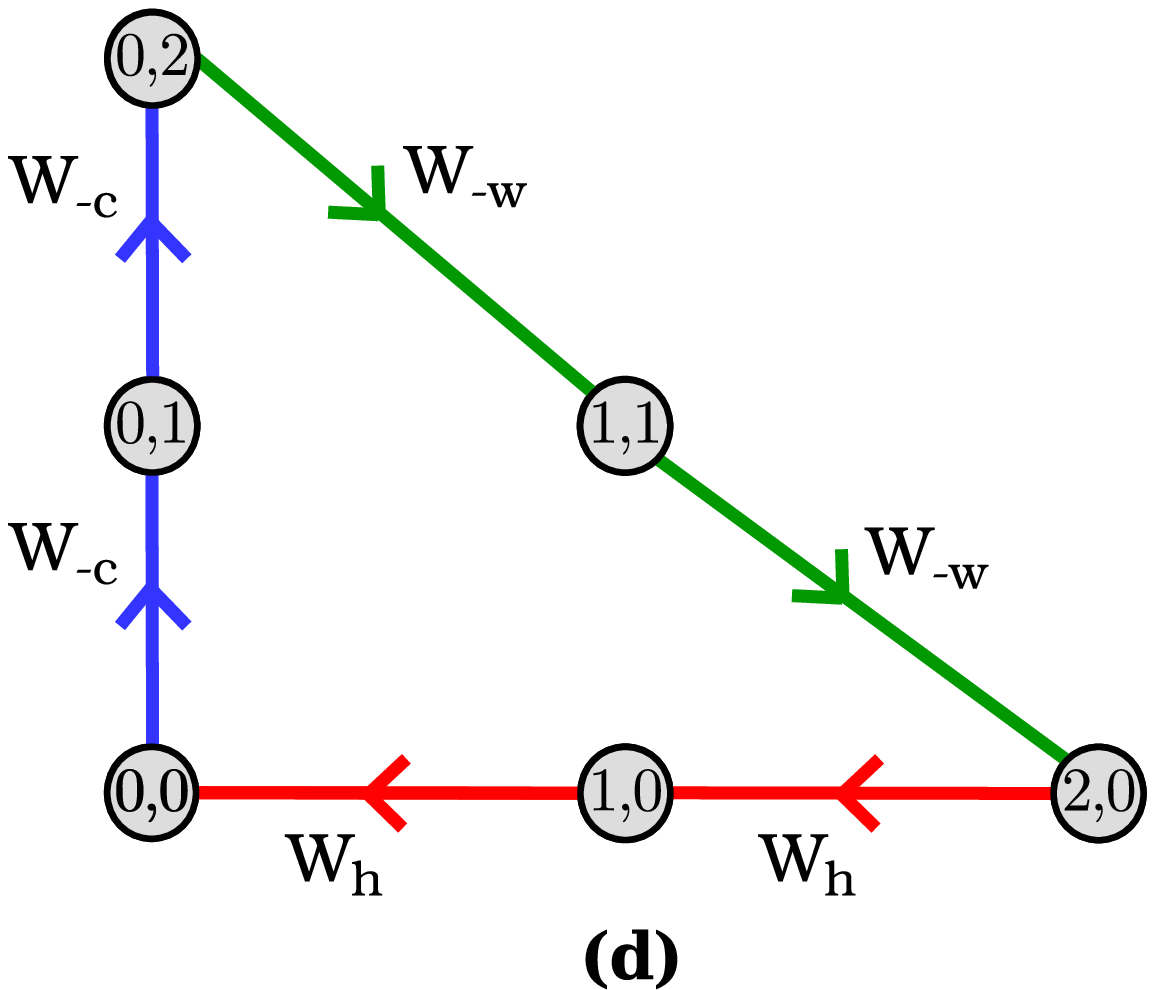}
\end{minipage}\hfill
\begin{minipage}{0.32\linewidth}
\includegraphics[width=\linewidth]{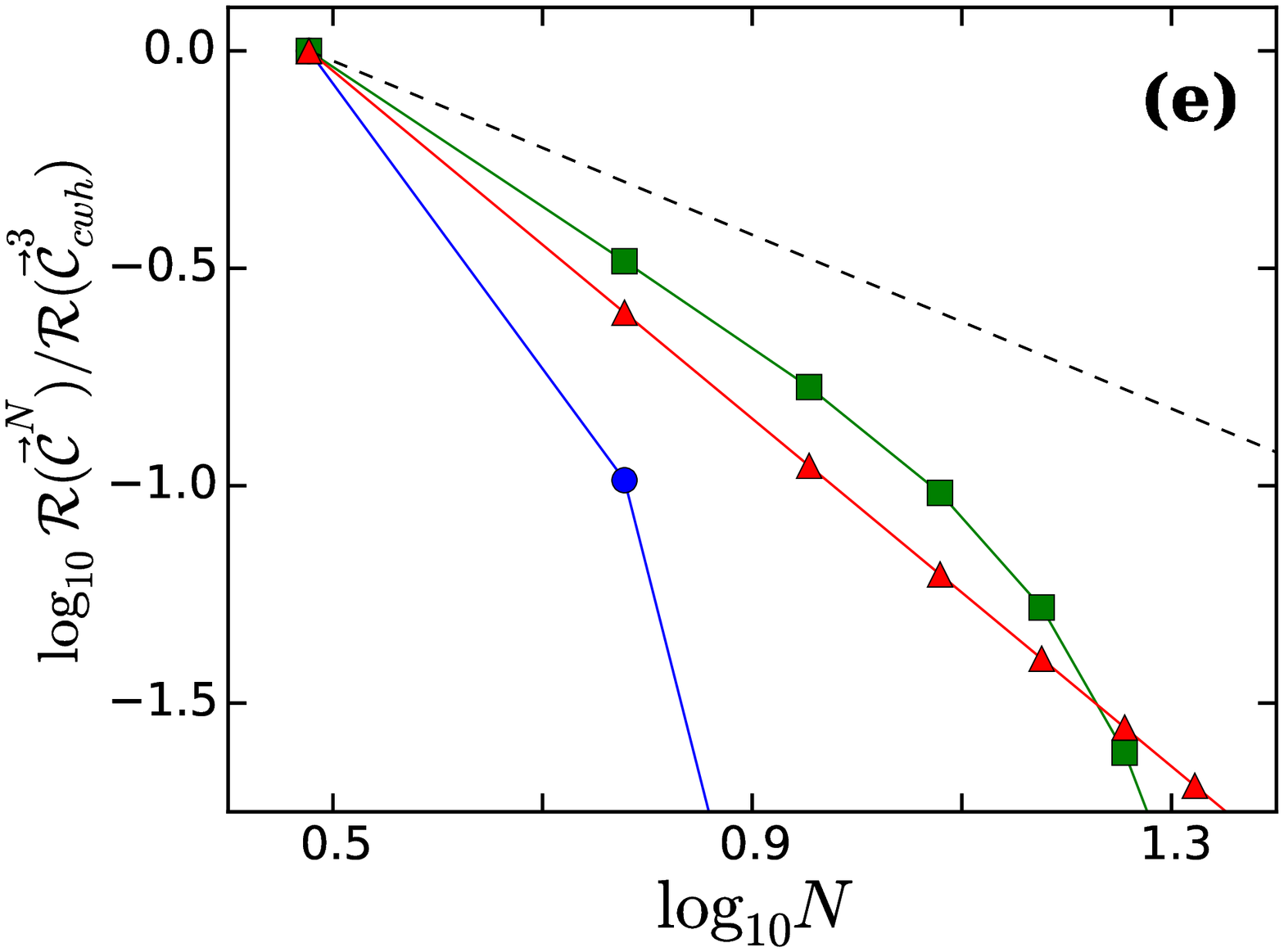}
\end{minipage}\hfill
\begin{minipage}{0.32\linewidth}
\includegraphics[width=\linewidth]{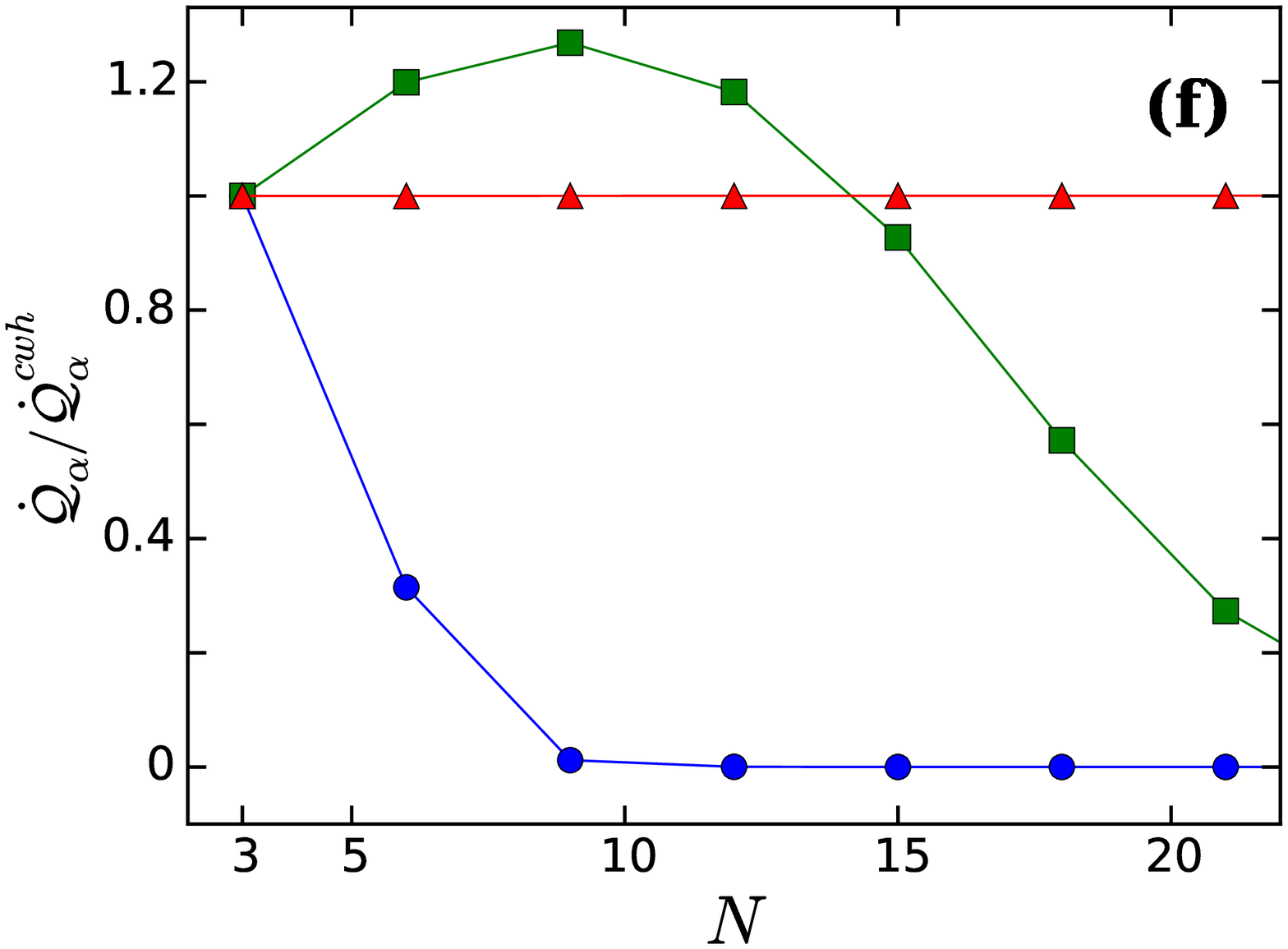}
\end{minipage}
\caption{Circuits (a) $\mathcal{C}^{3+2m_h}$ for $m_h=1$ and (d) $\mathcal{C}^{3+3n_+}$ for $n_+=1$. The triangle ($n_+=0$) is denoted by $\mathcal{C}^3_{cwh}$. The factor $\mathcal{R}(\vec{\mathcal{C}}^{\,N})$ and the heat currents (both normalized to the triangle values) as functions of the number of states $N$ are shown in (b) and (c) for $\mathcal{C}^{3+2m_h}$; (e) and (f) for $\mathcal{C}^{3+3n_+}$. The dashed lines follow a dependence $N^{-1}$. The calculations are performed using quantum systems described by the Hamiltonians and coupling operators given in \ref{sec:appendixHamiltonian}. The bath temperatures are parameterized by $t$, with  $t=0.3$ (circles), $t=1$  (squares) and $t=6000$ (triangles) corresponding to low, intermediate and high temperatures. The transition rates $W_{\pm \alpha}$ are calculated using (\ref{eq:ratesapp}) and (\ref{eq:Gammaapp}) with $d_\alpha=3$, $\gamma_c=\gamma_h=\gamma_w$,  $\omega_h=7$, $\omega_c=0.5$, $T_c=4t$, $T_h=5t$ and $T_w=6t$, in units for which $\hbar=k_B=\omega_0=1$. The lines are merely eye guides.}\label{fig:fig4}
\end{figure}

Considering these points, any circuit must be constructed adding either two-edge sets $\{\alpha\alpha\}$, with $\alpha=c,w,h$, or three-edge sets $\{cwh\}$ to guarantee that the change of energy of the system in a complete cycle is zero. We denote by $m=m_c+m_w+m_h$ the number of two-edge sets in a circuit. Each one of them contributes with the product $W_{-\alpha}W_{\alpha}$ to the algebraic value $\mathcal{A}(\vec{\mathcal{C}}^N)$, independently of the circuit orientation.  Circuits constructed only by adding two-edge sets ($N=2m$) are trivial circuits, $X^\alpha(\vec{\mathcal{C}}^{\;2m})=0$. The $n=n_+ + n_-$ three-edge sets $\{cwh\}$ in a circuit contribute either with the product $W_{-c}W_{-w}W_{h}$ (sets $n_+$) or $W_{c}W_{w}W_{-h}$ (sets $n_-$) to $\mathcal{A}(\vec{\mathcal{C}}^{\,N})$. Notice that when changing the circuit orientation to $-\vec{\mathcal{C}}$, $n_+$ and $n_-$ are interchanged. The smallest useful circuit is a triangle denoted by $\mathcal{C}^3$, see for example figure \ref{fig:fig1}(d) and (e). Large circuits $\mathcal{C}^N$ with $N=3n+2m$ states are obtained adding additional two and three-edges sets to $\mathcal{C}^3$. Their smallest instances are shown in figure \ref{fig:fig4}(a) and (d). 

\subsubsection{Circuit affinities and $\mathcal{R}(\vec{\mathcal{C}}^{\,N})$}

The circuit affinities are given by $X^\alpha(\vec{\mathcal{C}}^{\,N})=(n_+-n_-)X^\alpha(\vec{\mathcal{C}}^{\,3})$ for a proper choice of the cycles. Both $\mathcal{C}^N$ and $\mathcal{C}^3$ have the same value of the parameter $x$, provided that $n_+-n_-\neq 0$, and then the performance of the circuit $\mathcal{C}^N$, given by (\ref{eq:COP}) or (\ref{eq:efficiency}), is necessarily equal to the performance of $\mathcal{C}^3$. In other words, for a fixed set of transition rates the circuit performance is independent of the number of edges.

The remaining question is whether larger circuits result in an increment of the magnitude of the heat currents with respect to $\mathcal{C}^3$. As $X^\alpha(\vec{\mathcal{C}}^{\,N})$ increases at most linearly with $N$, the term depending on the affinities in (\ref{eq:onecircuitheatcurrent}) increases at most as $N^2$, but only when $NX(\vec{\mathcal{C}}^{\,3})/k_B$ remains small. However, the increment of the affinities with the number of states is compensated by the factor $\mathcal{R}(\vec{\mathcal{C}}^{\,N})$. As the number of terms in $D$ grows quadratically with $N$, one would expect that in most cases $\mathcal{R}(\vec{\mathcal{C}}^{\,N})$ decreases when adding new states and edges to the circuit. In fact, numerical evidence indicates that when adding two and three-edge sets to a circuit, $\mathcal{R}(\vec{\mathcal{C}}^{\,N})$ decreases equal or faster than $N^{-z}$, with $z\geq 1$ for large enough $N$, see for example figures \ref{fig:fig4}(b) and (e). Notice that we do not claim that $\mathcal{R}(\vec{\mathcal{C}}^{\,N'})<\mathcal{R}(\vec{\mathcal{C}}^{\,N})$ for arbitrary values of $N$ and $N'$ subjected to the condition $N'>N$. Our statement only applies to the construction where the circuit $\mathcal{C}^{\,N'}$ is obtained by adding two-edge and three-edge sets to $\mathcal{C}^{3}$, while keeping the edges and the orientation of the latter. Explicit expressions for $\mathcal{R}(\vec{\mathcal{C}}^{\,N})$ in the high and low temperature limits supporting this result are given in \ref{sec:appendixR}.

We have shown that typically the affinity term in (\ref{eq:onecircuitheatcurrent}) depends linearly on $N$ whereas $\mathcal{R}(\vec{\mathcal{C}}^{\,N})$ decreases faster than $N^{-1}$, and therefore in most cases the heat currents will decrease when adding additional edges to $\mathcal{C}^3$, see figures \ref{fig:fig4}(c) and (f). An increment in the  heat currents may be obtained at some extent by adding some three-edge sets when $z<2$ while the circuit affinity remains small enough to grow quadratically with the number of states, as shown in figure \ref{fig:fig4}(f) for intermediate temperatures. This improvement, although modest, may be relevant in situations where the heat currents are intrinsically small. Intuitively, increasing the circuit size implies the addition of states with larger energies and small populations except for specific values of the parameters. This small population makes harder closing the cycles and then effectively reduces the heat currents. Thus, the triangle $\mathcal{C}^3$ is in general the optimal choice as building block for multilevel devices.

\subsection{The triangle $\mathcal{C}^3$}

There are only two possible configurations of the triangle $\mathcal{C}^3$ compatible with condition (\ref{eq:conditionaffinity}): $\mathcal{C}^3_{cwh}$, shown in figure \ref{fig:fig1}(d), and $\mathcal{C}^3_{wch}$, where the cold and work edges are interchanged. This machine is one of the reference models used in quantum thermodynamics and it has been studied in both the $cwh$ \cite{Kosloff2014,Palao2001,Linden2010} and $wch$ \cite{Esposito2009} configurations. Since $X^\alpha(\vec{\mathcal{C}}^{\,3}_{cwh})=X^\alpha(\vec{\mathcal{C}}^{\,3}_ {wch})$ for a proper orientation, the circuits show the same thermodynamic performance. Notice that $\mathcal{A}(\vec{\mathcal{C}}^{\,3}_{cwh})=\mathcal{A}(\vec{\mathcal{C}}^{\,3}_{wch})$ but $D(\mathcal{C}^3_{cwh})\neq D(\mathcal{C}^3_{wch})$. Using (\ref{eq:onecircuitheatcurrent}), the heat currents are related by
\begin{equation}
\frac{\dot{\mathcal{Q}}_\alpha^{wch}}{\dot{\mathcal{Q}}_\alpha^{cwh}}=
\frac{D(\mathcal{C}^3_{cwh})}{D(\mathcal{C}^3_{wch})}\,.
\end{equation}
For high temperatures, $y_\alpha\equiv\exp[-E_\alpha/(k_BT_\alpha)]\approx 1$, the arrangement of the edges in the circuit is irrelevant and $\dot{\mathcal{Q}}_\alpha^{wch}/\dot{\mathcal{Q}}_\alpha^{cwh}\approx 1$. A different picture appears at low temperatures, $y_\alpha\ll 1$,
\begin{equation}
\frac{\dot{\mathcal{Q}}_\alpha^{wch}}{\dot{\mathcal{Q}}_\alpha^{cwh}}\approx
\frac{W_{c}W_{h}+W_{c}W_{w}}{W_{c}W_{w}+W_{w}W_{h}}\,.
\end{equation}
For $W_{c}<W_{w}$, the ratio $\dot{\mathcal{Q}}_\alpha^{wch}/\dot{\mathcal{Q}}_\alpha^{cwh}<1$ and for $W_{c}>W_{w}$, $\dot{\mathcal{Q}}_\alpha^{wch}/\dot{\mathcal{Q}}_\alpha^{cwh}>1$. Then the most favorable configuration corresponds to the lowest transition rate being associated with transitions from the ground state, by far the most populated in the low temperature limit.

\section{Thermal machines represented by a graph with multiple circuits}\label{sec:multiplecircuits}

We study now multilevel absorption machines with multiple circuits. We start  by analyzing the relation between the heat currents and the performance of a circuit $\mathcal{C}_\nu$ in an arbitrary graph $\mathcal{G}$, and the corresponding quantities for the (isolated) graph circuit $\mathcal{C}_\nu^{iso}$. To this end, we rewrite (\ref{eq:cycleheatcurrent}) as
\begin{equation}\label{eq:multipleheatcurrent}
\dot{q}_\alpha(\mathcal{C}_\nu)=D(\mathcal{G})^{-1}\det(-\mathbf{W}|\mathcal{C}_\nu)D(\mathcal{C}_\nu^{iso})\,\dot{q}_\alpha(\mathcal{C}_\nu^{iso})\,.
\end{equation}
Using this expression we find: 

\begin{itemize}
\item[(i)] $|\dot{q}_\alpha(\mathcal{C}_\nu)| < |\dot{q}_\alpha(\mathcal{C}_\nu^{iso})|$.

\item[(ii)] $\varepsilon(\mathcal{C}_\nu)=\dot{q}_c(\mathcal{C}_\nu)/\dot{q}_w(\mathcal{C}_\nu)=\varepsilon(\mathcal{C}_\nu^{iso})$ and $\eta(\mathcal{C}_\nu)=-\dot{q}_w(\mathcal{C}_\nu)/\dot{q}_h(\mathcal{C}_\nu)=\eta(\mathcal{C}_\nu^{iso})$. 
\end{itemize}

The first result indicates that the magnitude of the heat currents associated with a circuit in a graph is always smaller than the one corresponding to the isolated circuit. It follows from (\ref{eq:determinant}) by noticing that the product between a term in the forest $\det(-\mathbf{W}|\mathcal{C}_\nu)$  and a term of $D(\mathcal{C}_\nu^{iso})$ gives the algebraic value of one of the oriented maximal trees of $\mathcal{G}$. Therefore $\det(-\mathbf{W}|\mathcal{C}_\nu)D(\mathcal{C}_\nu^{iso})=\sum_{\mu=1}^{N_T^\prime}\sum_{i\in \nu}\mathcal{A}(\vec{\mathcal{T}}_i^\mu)$, with $\sum_{i\in \nu}$ the summation over all the vertices of $\mathcal{C}_\nu$ and being the number of maximal trees involved $N_T^\prime\leq N_T$ . The second result derives directly from (\ref{eq:COP}) and (\ref{eq:efficiency}) and indicates that the circuit performance is not modified when the circuit is included in an arbitrary graph.

\subsection{General bound for the performance}

A consequence of (ii) is that the device performance cannot exceed the corresponding to the circuit with the best performance. For example, let us consider a device working as an absorption refrigerator, $\dot{\mathcal{Q}}_c$ and $\dot{\mathcal{Q}}_w>0$. The coefficient of performance is given by
\begin{equation}
\varepsilon = 
\sum_{\nu=1}^{N_C^\prime}\frac{\dot{q}_w(\mathcal{C}_\nu)}{\dot{\mathcal{Q}}_w}
\varepsilon(\mathcal{C}_\nu) - 
\sum_{\nu=N_C^\prime+1}^{N_C^{\prime\prime}}\frac{|\dot{q}_c(\mathcal{C}_\nu)|}{\dot{\mathcal{Q}}_w}\,,
\end{equation}
where $\dot{q}_c(\mathcal{C}_\nu)$ is positive for the $N_C^\prime$ circuits contributing to the cooling cycle, and negative for the $N_C^{\prime\prime}-N_C^\prime$ ``counter-contributing'' circuits, corresponding for example to heat leaks and circuits with finite counter-currents which flow in directions against the operation mode \cite{Correa2015,Gonzalez2016}. The $N_C-N_C^{\prime\prime}$ trivial circuits are irrelevant in this discussion. In consequence, denoting by $\varepsilon(\mathcal{C}_\nu)_{max}$ the largest performance of a circuit in the graph,
\begin{equation}
\varepsilon\leq \varepsilon(\mathcal{C}_\nu)_{max}\,,
\end{equation}
and the equality, $\varepsilon=\varepsilon(\mathcal{C}_\nu)_{max}$, is reached when $N_C^{\prime\prime}-N_C^\prime=0$ and $\varepsilon(\mathcal{C}_\nu)=\varepsilon(\mathcal{C}_\nu)_{max}$ for all the circuits. In particular, $\varepsilon=\varepsilon_C$ only if all of them achieve the Carnot performance for the same value of the affinity. A similar analysis applies to the device working as a heat transformer. Therefore, with regard to the performance, optimal multilevel machines are represented by graphs without ``counter-contributing'' circuits. We will impose this condition in the design of the optimal graph.

\subsection{Graph topology and heat currents} 

The magnitude of the physical heat currents is determined by the graph topology and the value of the transition rates. We first explore the graph topology of an arbitrary graph with the only restriction that two vertices can be connected by just one edge (see section \ref{sec:subcircuits}). 

In general $\dot{\mathcal{Q}}_\alpha=\sum_{\nu=1}^{N_C}\dot{q}_\alpha(\mathcal{C}_\nu)$ increases with the number of positive contributing circuits $N_C^\prime\leq N_C$, which operate in the same way as the entire device. However, this increment may be hindered by the unavoidable decrease in $D(\mathcal{G})^{-1}\det(-\mathbf{W}|\mathcal{C}_\nu)$ when adding new states and edges to a graph. The factor $D$ is the sum of $NN_T$ terms. For circuits of length $L$, $\det(-\mathbf{W}|\mathcal{C}_\nu^L)$ is the sum of $\det(\widetilde{\mathbf{A}}|\mathcal{C}_\nu^L)$ terms. In this expression the submatrix $(\widetilde{\mathbf{A}}|\mathcal{C}_\nu^L)$ is obtained by removing from  $\widetilde{\mathbf{A}}$ all the rows and columns corresponding to the vertices of the circuit $\mathcal{C}_\nu^L$. The matrix $\widetilde{\mathbf{A}}$ is calculated by replacing the diagonal elements $a_{ii}$ of the adjacency matrix $\mathbf{A}$ (see for example \cite{Foulds1992}) by the vertex degree of the corresponding state $i$. The non diagonal elements are $a_{ij}=1$ when states $i$ and $j$ are connected by an edge, and $a_{ij}=0$ otherwise. Therefore, when attending to the number of terms, the magnitude of the heat currents resulting from the positive contributions of $N_L\leq N_C^\prime$ circuits of length $L$ is related to the topological parameter
\begin{equation}
\tau_L\equiv \frac{1}{N}\sum_{\nu=1}^{N_L}\lambda(\mathcal{C}_\nu^L)
<\tau_L^b \equiv \frac{N_L}{N}\,,
\end{equation}
where $\lambda(\mathcal{C}_\nu^L)\equiv\det(\widetilde{\mathbf{A}}|\mathcal{C}_\nu^L)/N_T$, with $N_T^{-1}\leq \lambda(\mathcal{C}_\nu^L)<1$. The ratio $\lambda$ may depend on the position of the circuit in the graph and in general $\lambda(\mathcal{C}_\nu^{L^\prime})<\lambda(\mathcal{C}_\nu^L)$ when $L^\prime>L$. Although $\tau_L$ can be readily calculated, we found that the upper bound $\tau_L^b$ incorporates the relevant information about the graph topology. In particular, it makes clear the relevance of the graph connectivity: favorable graphs consist in as many small positive contributing circuits as possible (that is, avoiding heat leaks and another negative contributions), built with the smallest possible number of states, implying a large graph connectivity. This dependence on the graph topology is weighted by the transition rates. For high temperatures all circuits participate in the heat currents. However, only small circuits including the ground state will contribute significantly in the low temperature regime, independently of the total number of circuits in the graph. 

\subsubsection{Graphs constructed by merging triangles}

The optimal choice for the building block is the triangle, the  smallest possible contributing circuit as described before. We consider that all the triangles have fixed energy gaps for transitions assisted by the same bath. This is a necessary condition to achieve the maximal possible connectivity because otherwise adjacent triangles cannot share any edge. In order to analyze the dependence on the graph topology we consider now the more restrictive condition of fixed transition rates for each bath. This assumption will be relaxed latter. Moreover, we assume the PCD condition, that implies now that the maximum vertex degree in the graph is six, i.e. each state may be connected at most to another six ones, see figure \ref{fig:fig3}(b). As a consequence, all the constructed graphs are planar and $\tau_3^b$ incorporates the relevant topological information. The number triangles is easily accessible by using the adjacency matrix, $N_3=\Tr\{\mathbf{A^3}\}/6$, where $\Tr\{\}$ denotes the trace.

\begin{figure}[h] 
\centering
\begin{minipage}{0.45\linewidth}
\includegraphics[width=\linewidth]{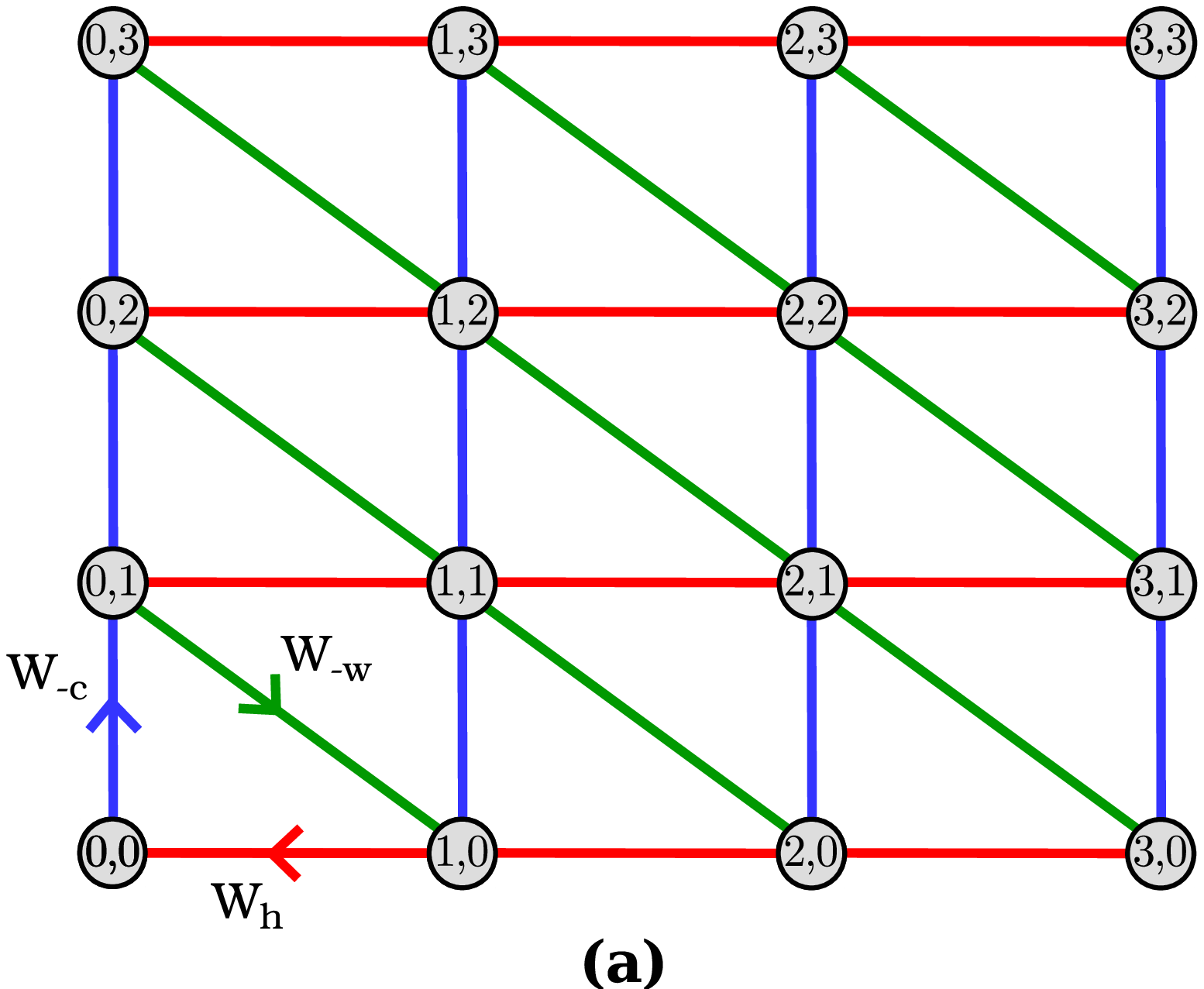}
\end{minipage}
\begin{minipage}{0.45\linewidth}
\includegraphics[width=\linewidth]{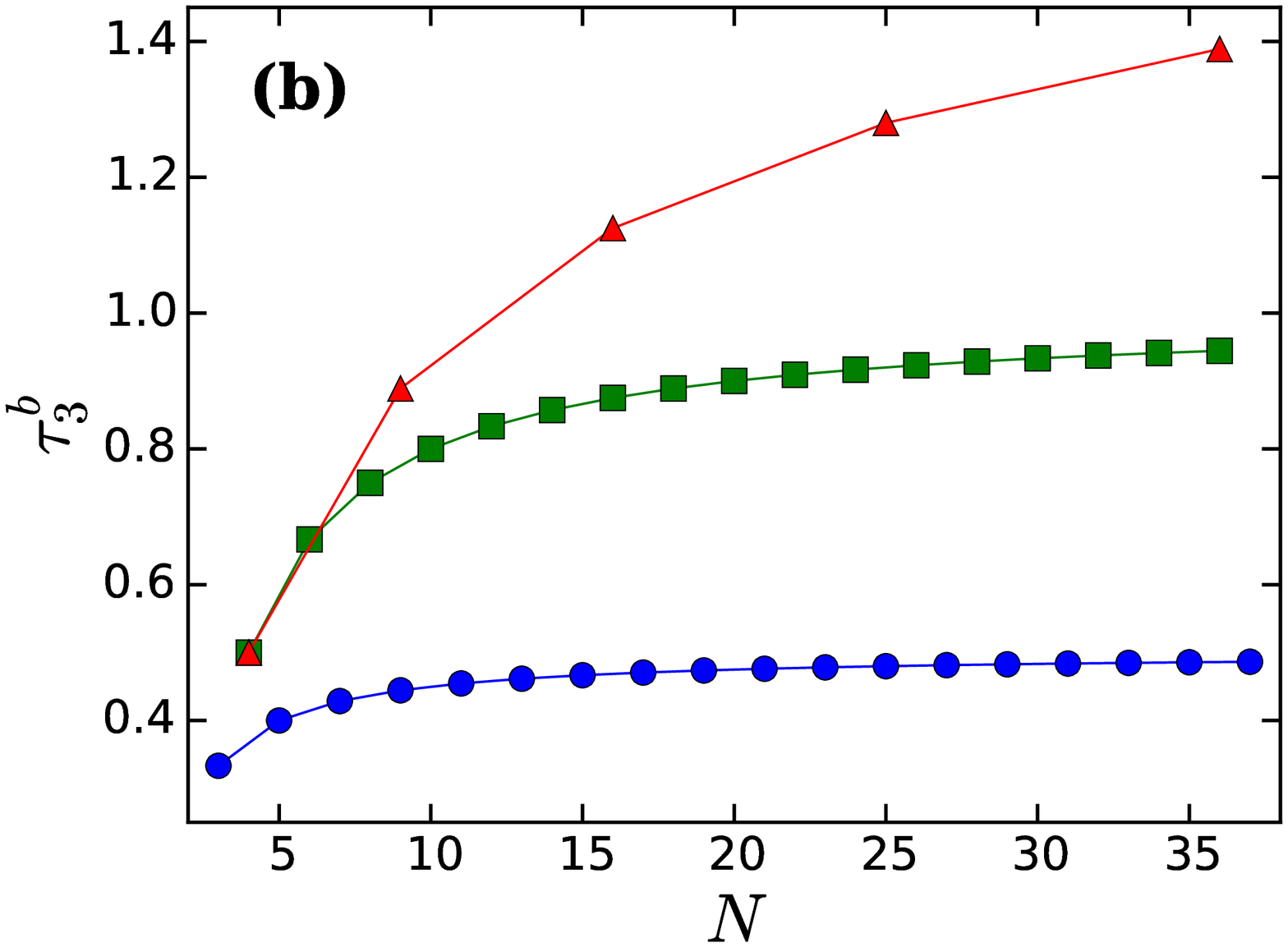}
\includegraphics[width=\linewidth]{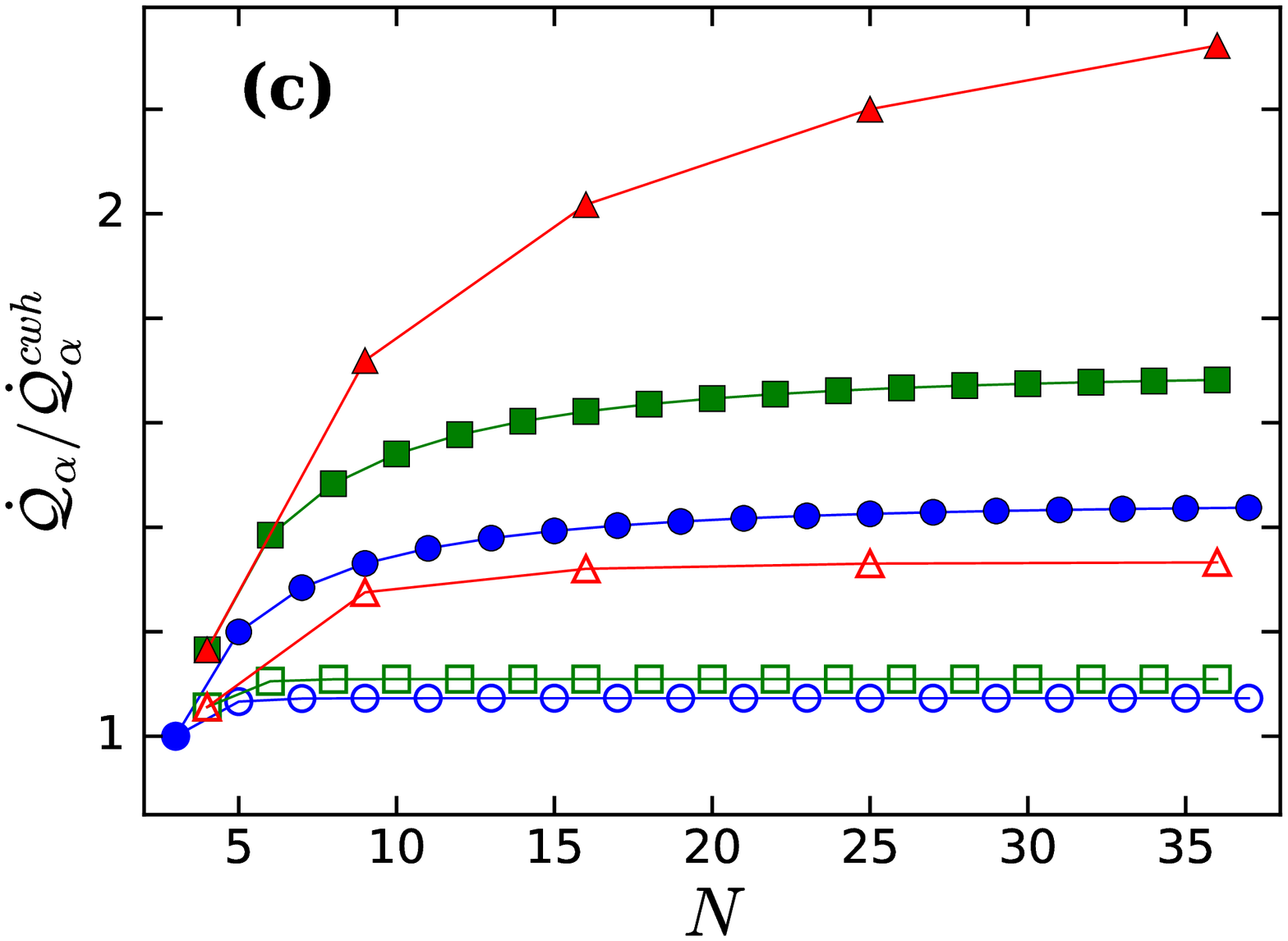}
\end{minipage}
\caption{(a)  Graph $\mathcal{G}_4^B$. States are labelled by the pair $(n_c,n_h)$, being the state energy $n_cE_c+n_hE_h$. (b) $\tau_3^b$ as a function of the number of states for $\mathcal{G}_4^B$ (triangles), $\mathcal{G}_{4L}^B$ (squares) and $\mathcal{G}_{3}^B$ (circles). (c) Heat currents (normalized to $\dot{\mathcal{Q}}_\alpha^{cwh}$) for two temperature regimes given by $t=0.07$ (empty symbols) and $t=1$ (solid symbols) where $T_c=5t$, $T_h=6t$ and $T_w=7t$. The remaning parameters are $d_\alpha=1$, $\gamma_c=\gamma_h=\gamma_w$,  $\omega_h=1$, $\omega_c=0.5$, in units for which $\hbar=k_B=\omega_0=1$. The calculations are performed for quantum systems described in \ref{sec:appendixHamiltonian}. The lines are merely eye guides.}\label{fig:fig5}
\end{figure}

The graph with the largest connectivity compatible with our restrictions is denoted by $\mathcal{G}_4^B$. It is constructed using $B$ units of two triangles sharing one edge, for example one associated with the work bath, see figure \ref{fig:fig5}(a). We consider square graphs with $1,4,9,\dots$ units, being the smallest instance $\mathcal{G}_4^{B=1}\equiv\mathcal{G}_4$. By construction, the two configurations of the triangle, $cwh$ and $wch$, are present. Besides, many other circuits can be identified. For example $\{(0,0),(0,1),(1,1),(2,0),(1,0),(0,0)\}$ is a circuit $\mathcal{C}^{3+2m_h}$ with $m_h=1$, and $\{(0,0),(0,1),(0,2),(1,1),(2,0),(1,0),(0,0)\}$ a circuit $\mathcal{C}^{3+3n_+}$ with $n_+=1$. All of them follow the same operation mode. There are also many trivial circuits, for example $\{(0,0),(0,1),(1,1),(1,0),(0,0)\}$. This construction is optimal with respect to the performance because it can attain the reversible limit as there are not ``counter-contributing'' circuit, see the discussion for $\mathcal{C}_3$ in section \ref{sec:motivation}.

We also consider two subgraphs of $\mathcal{G}_4^B$ for comparison purposes. The first one is a row of this units, denoted by $\mathcal{G}_{4L}^B$, which represents the absorption device studied in \cite{Correa2014c}. The second one is obtained considering only a row and removing the upper hot edges. We use this graph, denoted by $\mathcal{G}_{3}^B$, to compare $\tau^b_L$ with other measure of the graph connectivity in \ref{sec:appendixG3B}.

Figure \ref{fig:fig5}(b) shows the parameter $\tau_3^b$ for $\mathcal{G}_4^B$, $\mathcal{G}_{4L}^B$ and $\mathcal{G}_3^B$, considering only complete units in each case. For a given number of states, larger values of $\tau_3^b$ correspond to larger number of circuits and therefore to a larger connectivity. When the number of states increases the parameter $\tau_3^b$ saturates to a different constant value in each case. This is reflected in the physical heat currents shown in figure \ref{fig:fig5}(c) for different bath temperatures. This saturation is due to the difficulty of exploring big circuits or those which are distant from the ground state in complex graphs. The simple picture based on the parameter $\tau_3^b$ is weighted by the transition rates. For decreasing bath temperatures, all the currents converge to the same result, independently of the number of circuits, since only the triangle including the ground state contributes significantly to them.

In summary, given a set of transition rates and a number of levels, the best topology corresponds to the most connected planar graph $\mathcal{G}_4^B$. This construction only contains trivial and positive contributing circuits and provides in general the largest heat currents for fixed rates.

\subsection{Transition rates and heat currents}

\begin{figure}[h]
\centering
\begin{minipage}{0.45\linewidth}
\includegraphics[width=\linewidth]{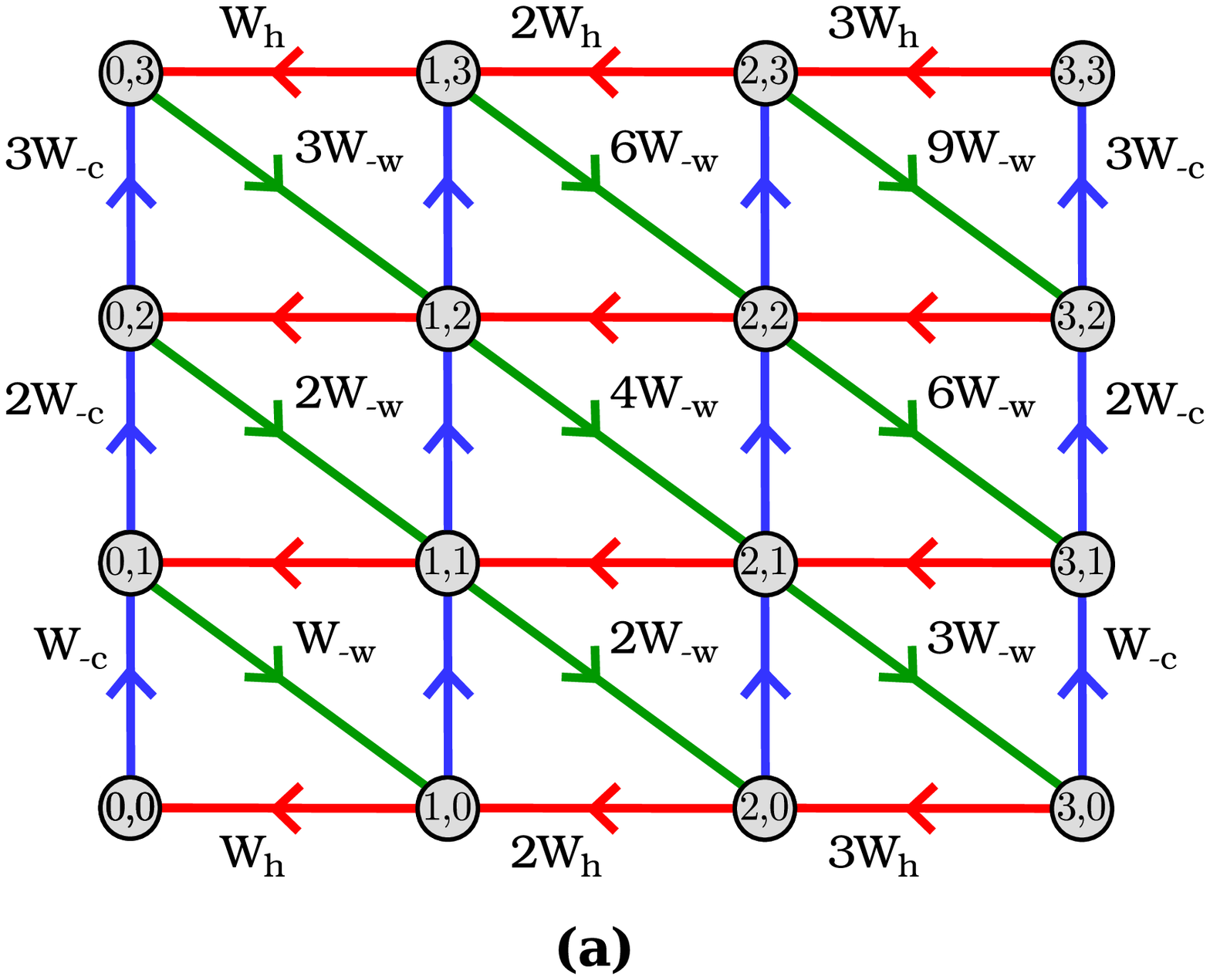}
\end{minipage}
\begin{minipage}{0.45\linewidth}
\includegraphics[width=\linewidth]{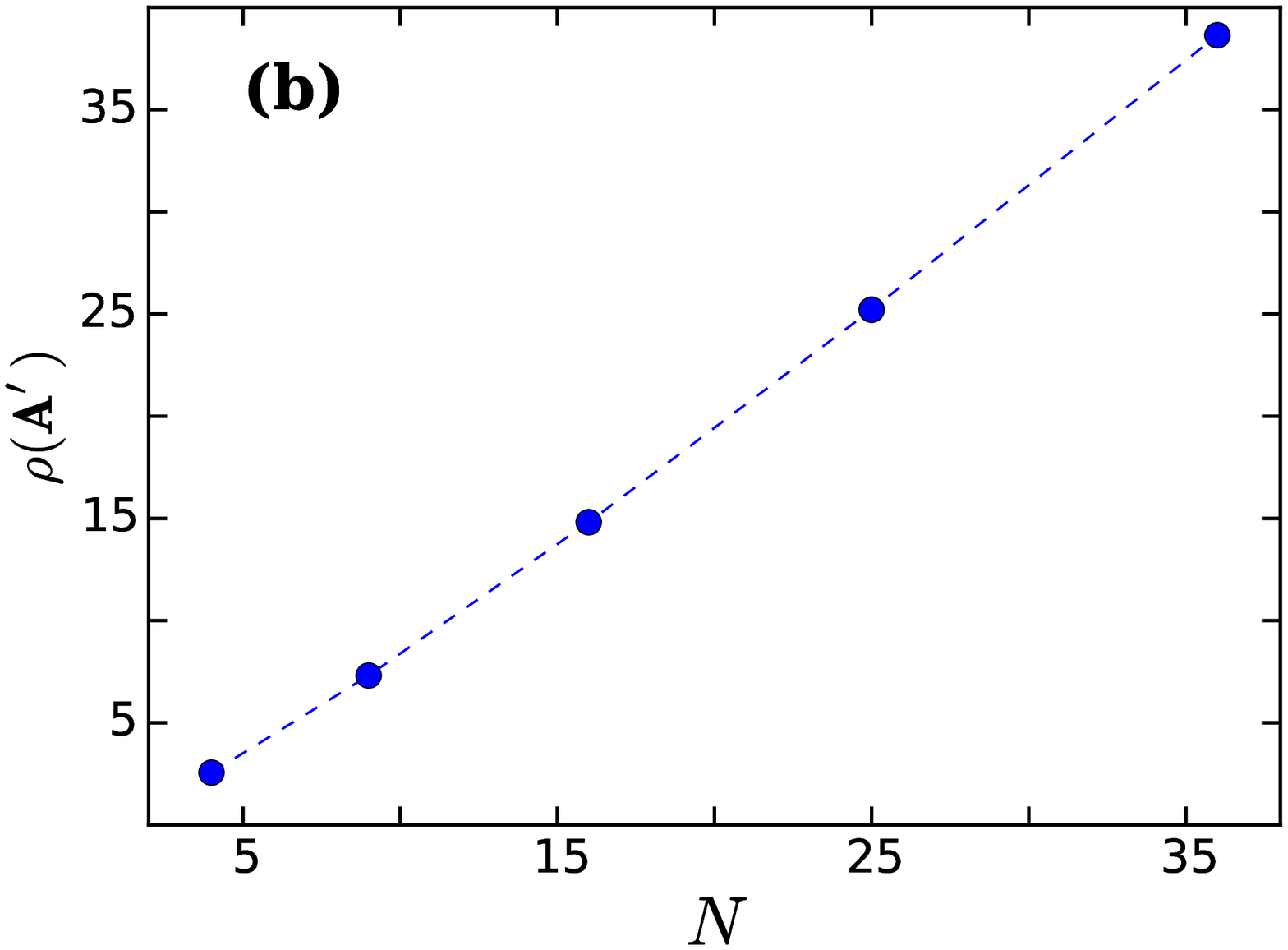}
\includegraphics[width=\linewidth]{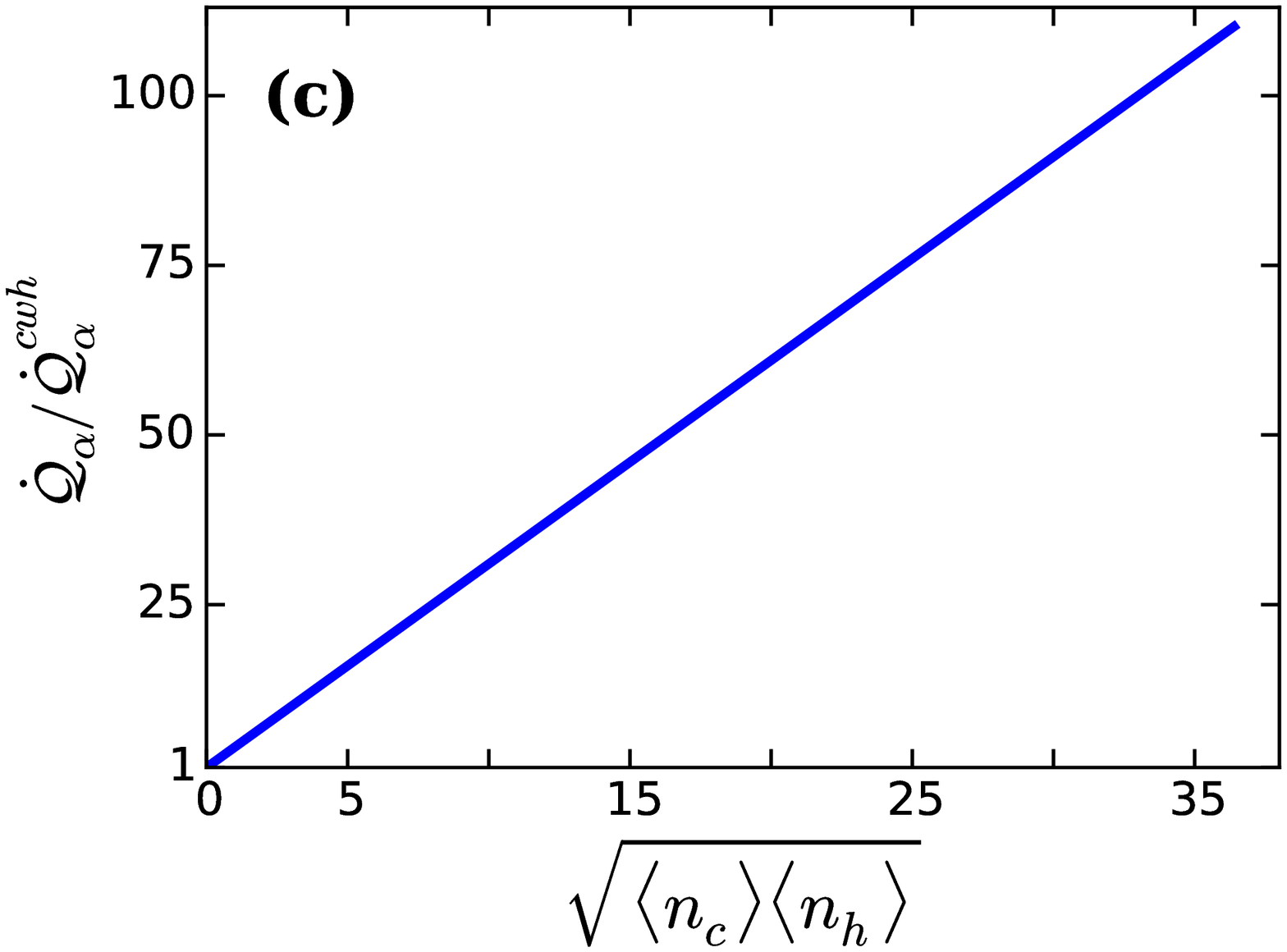}
\end{minipage}
\caption{(a) The graph $\mathcal{G}_{HO}^B$. (b) Spectral radius as a function of the number of states. The line is merely an eye guide. (c) Heat currents (normalized by $\dot{\mathcal{Q}}_\alpha^{cwh}$) as a function of $\sqrt{\langle n_c\rangle \langle n_h \rangle}$, calculated for different bath temperatures parametrized by $t$, with $d_\alpha=3$, $\gamma_c=\gamma_h=\gamma_w$, $\omega_h=1$, $\omega_c=0.5$, $T_c=30t$, $T_h=34t$ and $T_w=10^6$, in units for which $\hbar=k_B=\omega_0=1$.}\label{fig:fig6}
\end{figure}

We have shown that for fixed transition rates the heat currents saturate to a constant value when increasing the number of states. To overcome this limitation, we now consider a graph with the optimal topology given by $\mathcal{G}_4^B$ and relax the condition on the rates but keeping fixed energy gaps. The circuit affinities and then the performance are not modified. Considering (\ref{eq:nondiag}), all the transition rates must be taken as $sW_{\pm\alpha}$, with $s\geq 1$, and $W_{\pm\alpha}$ the smallest rate. As discussed for circuit graphs, increasing $s$ will lead to larger heat currents.

In particular, we analyze the construction shown in figure \ref{fig:fig6}(a), denoted by $\mathcal{G}_{HO}^B$, which has a simple physical implementation as discussed below. Seeking a measure of the graph connectivity when the transition rates increase with $s$, and in analogy with the adjacency matrix, we define $\mathbf{A}^\prime$ with elements $a_{ij}^\prime=s$ when states $i$ and $j$ are adjacent with transition rates $sW_{\pm\alpha}$, and we denote its spectral radius as $\rho(\mathbf{A}^\prime)$, see \ref{sec:appendixG3B}. When incorporating additional building units into the graph, the spectral radius defined in this way increases nearly linearly with the number of states, see figure \ref{fig:fig6}(b). 

The graph $\mathcal{G}_{HO}^B$, allowing an infinity number of building blocks, represents the master equation of a device composed of two harmonic oscillators \cite{Levy2012}. Each oscillator is connected to a thermal bath at temperatures $T_c$ and $T_h$. The coupling operators are $\hat{S}_c=\hat{a}_c$ and  $\hat{S}_h=\hat{a}_h$ (see \ref{sec:appendixrates}), being $\hat{a}_\alpha$ the annihilation operator of the oscillator coupled to the bath $\alpha$. A third bath at temperature $T_w$ is coupled to the system through the operator $\hat{S}_w=\hat{a}_c^\dagger\hat{a}_h$. For simplicity we assume a very large value of $T_w$, a regime for which the heat currents can be easily calculated. Figure \ref{fig:fig6}(c) shows the heat currents as a function of $\sqrt{\langle n_c\rangle \langle n_h \rangle}$, which gives a rough estimation of the number of states populated and then of the effective graph size, calculated for increasing bath temperatures. In this expression $\langle n_\alpha\rangle$ is the average number of excitations in the oscillator $\alpha=c,h$. When the temperature increases, larger areas of the graph are populated involving a larger number of circuits, which results in an increment of the magnitude of the heat currents. This example illustrates that given a machine with the optimal topology, the rates can always be carefully designed to achieve larger currents without diminishing the performance.

\section{Conclusions}\label{sec:conclusion}

We have determined the steady state heat currents associated with all possible circuits in the graph representing the master equation of multilevel continuous absorption machines. Each circuit is related to a thermodynamically consistent mechanism in the device functioning. Although the number of circuits may be very large when increasingly complex graphs are considered, efficient standard algorithms, which scale as $N_C(N+2U)$ \cite{Johnson1975}, can be used for determining them. For example, in the graphs studied in previous sections $U$ increases linearly and $N_C$ quadratically with the number of states and the computational cost scales as $N^3$. The main result of the decomposition is an equation for the circuit heat currents depending only on the transition rates, without any prior knowledge of the steady state populations. This expression allows us to analyze the two relevant quantities for refrigerators and heat transformer, the magnitude of the physical heat currents and the performance. We focus on devices coupled to three baths, since they can provide the same currents than more complicated setups.

In order to elucidate the role of the graph topology in the thermodynamic properties, we have analyzed machines constructed by a fixed set of transition rates. In devices represented by a single graph circuit, the performance depends only on the circuit affinities, which can be tuned to reach the reversible limit, and the magnitude of the heat currents decreases in general with the number of states. Then the simplest graph, a triangle, leads to the largest heat currents in most cases and is the proper building block for optimal multilevel machines.

When considering generic devices, we have found that the performance of the device cannot exceed the corresponding to the circuit with maximum performance. Besides the magnitude of the heat currents is described by a topological parameter that increases with the graph connectivity. As a consequence, if the construction of larger graphs including additional circuits presents a limited connectivity, then the magnitude of the resulting physical heat currents saturates to a constant value, which is different for different constructions. We use triangles with fixed energy gaps for transitions assisted by the same bath to construct the graph with the largest possible connectivity, denoted by  $\mathcal{G}_4^B$. This is a planar graph containing neither heat leaks nor ``counter-contributing'' circuits.

The assumption of a fixed set of transition rates can be relaxed. We give the necessary condition to improve the currents without modifying the performance. We provide an example using a system of harmonic oscillators. In this case the magnitude of the heat currents increases almost linearly with the effective size of the graph, determined by the achievable range of temperatures. An interesting question is whether there are other physical feasible implementations leading to a faster than linear dependence of the currents on the number of states.

The circuit decomposition could be employed in other different scenarios, from the study of heat transport through quantum wires to the analysis of machines designed for complicated tasks involving more than three baths. Besides, our formalism also applies to the case of reservoirs exchanging both energy and particles with the system, and even to periodically driven machines. The only condition required is that the population and coherence dynamics are decoupled in a certain basis. However, this is not always possible, as for example in weakly driven systems. Finally, it is worth to mention that the study of four-stroke many-particle thermal machines has been recently addressed in \cite{delcampo2016}, the analysis of their continuous counterparts is another interesting issue we can explore in the future by using the circuit decomposition. We expect these findings will help in the experimental design of absorption devices.

\ack 

We thank L. Correa and A. Ruiz for useful discussions. J. Onam Gonz{\'a}lez acknowledges a Formaci{\'o}n de Profesorado Univeritario (FPU) fellowship from the Spanish Ministerio de Educaci{\'o}n, Cultura y Deportes (MECD). Financial support by the Spanish Ministerio de Econom{\'i}a y Competitividad (MINECO) (FIS2013-41352-P) and European Cooperation in Science and Technology (COST) Action MP1209 is gratefully acknowledged.

\appendix


\section{Transition rates for quantum systems weakly coupled with thermal baths}
\label{sec:appendixrates}

In this appendix we describe how to calculate the transition rates $W_{ij}^\alpha$ in the master equation (\ref{eq:master}) for a quantum system with Hamiltonian $\hat{H}_S=\sum_{i=1}^N\hbar\omega_i|i\rangle\langle i|$, and coupled with $R$ bosonic baths at temperatures $T_\alpha$. We assume that the PCD condition holds. The total Hamiltonian reads
\begin{equation}\label{eq:totalhamiltonian}
\hat{H}\,=\,\hat{H}_{S}\,+\,
\sum_{\alpha=1}^R\,\left(\hat{H}_{S,\alpha}\,+\,\hat{H}_\alpha\right)\,,
\end{equation}
where $\hat{H}_\alpha$ are the bath Hamiltonians and the coupling terms are given by
\begin{equation}\label{eq:couplingterm}
\hat{H}_{S,\alpha}\,=\,\hbar\sqrt{\gamma_\alpha}(\hat{S}_\alpha+\hat{S}_\alpha^\dagger)\otimes\hat{B}_\alpha\,,
\end{equation}
with $\hat{S}_\alpha$ and $\hat{B}_\alpha$ a system and a bath operator respectively. The rates $\gamma_\alpha$ determine the time scale of the system relaxation dynamics. Finally, the system operators in the coupling terms are 
\begin{equation}\label{eq:couplingoperators}
\hat{S}_\alpha\,=\,\sum_{i=1}^N \sum_{j>i} \, c_{ij}^\alpha\,\,|i\rangle\langle j|\,.
\end{equation}
We consider the following assumptions: the system is weakly coupled with the environments, $\hbar\gamma_\alpha\ll k_B T_\alpha$, and $\gamma_\alpha\ll|\omega_{ij}-\omega_{i'j'}|$, with $\omega_{ij}\neq \omega_{i'j'}$ and $\omega_{ij}=\omega_j-\omega_i$. Then the Born-Markov and the rotating wave approximation applies and the master equation for the populations of the eigenstates of $\hat{H}_S$ \cite{Breuer2002} is given by (\ref{eq:master}) with transition rates ($i<j$) 
\begin{equation}\label{eq:ratesapp}
W_{ij}^\alpha\,=\,\gamma_\alpha\,|c_{ij}^\alpha|^2\,\Gamma^\alpha_{\omega_{ij}}\,.
\end{equation}
The functions $\Gamma^\alpha$ only depend on bath operators 
\begin{equation}\label{eq:Gammaapp}
\Gamma^\alpha_{\omega}\,=\,2\Re\left\{
\int_0^\infty \,dt \exp(\mathbf{i}\, \omega t)\,{\rm Tr}_\alpha[
\hat{B}_\alpha(t) \hat{B}_\alpha \hat{\rho}_\alpha] \right\}\,,
\end{equation}
where $\hat{B}_\alpha(t)=\exp(\mathbf{i}\,\hat{H}_\alpha t/\hbar) \hat{B}_\alpha \exp(-\mathbf{i}\,\hat{H}_\alpha t/\hbar)$ and $\hat{\rho}_\alpha$ denotes the bath thermal state. We will consider bosonic baths of physical dimensions $d_\alpha$ and coupling operators $\hat{B}_\alpha\propto\sum_\mu\sqrt{\omega_\mu}(\hat{b}_\mu^\alpha+\hat{b}_\mu^{\alpha\dagger})$. The summation is over all the bath modes of frequencies $\omega_\mu$ and annhilation operators $\hat{b}_\mu$. With this choice the rates $\Gamma^\alpha_{\pm\omega}$ are \cite{Breuer2002}
\begin{eqnarray}\label{eq:rates}
\Gamma^\alpha_{\omega} &=& \,(\omega/\omega_0)^{d_\alpha}[N^\alpha(\omega)+1] \,,
\nonumber \\
\Gamma^\alpha_{-\omega} &=& \Gamma^\alpha_{\omega}\exp(-\omega\hbar/k_BT_\alpha) \,,
\end{eqnarray}
with $N^\alpha(\omega)=[\exp(\omega\hbar/k_BT_\alpha)-1]^{-1}\,$. The frequency $\omega_0$ depends on the physical realization of the coupling with the bath. The condition (\ref{eq:diag}) derives now directly from the conservation of the normalization of the system density matrix. Besides, the Kubo-Martin-Schwinger relation in (\ref{eq:rates}) implies (\ref{eq:nondiag}).


\section{Quantum implementation of the graphs}
\label{sec:appendixHamiltonian}

We introduce here a possible quantum physical realization of the graphs described in the main text by specifying their Hamiltonians and coupling operators. Considering bosonic heat baths, the results of \ref{sec:appendixrates} can be used to obtain the corresponding transition rates. In all cases $\omega_c+\omega_w=\omega_h$.

\begin{itemize}
\item[(i)] $\mathcal{G}_4$.
\begin{equation}
\hat{H}_S=\hbar[\omega_c|2\rangle\langle 2|+\omega_h|3\rangle\langle 3|+(\omega_h+\omega_c^\prime)|4\rangle\langle 4|]\,,
\end{equation}
and $\hat{S}_c=|1\rangle\langle 2|+|3\rangle\langle 4|$, $\hat{S}_w=|2\rangle\langle 3|$, $\hat{S}_h=|1\rangle\langle 3|+|2\rangle\langle 4|$. The two-qubit model \cite{Levy2012} corresponds to $\omega_c^\prime=\omega_c$.

\item[(ii)] $\mathcal{C}^3_{cwh}$.
\begin{equation}
\hat{H}_S=\hbar(\omega_c\,|2\rangle\langle 2|+\omega_h\,|3\rangle\langle 3|)\,,
\end{equation}
and $\hat{S}_c=|1\rangle\langle 2|$, $\hat{S}_w=|2\rangle\langle 3|$, $\hat{S}_h=|1\rangle\langle 3|$.

\item[(iii)] $\mathcal{C}^3_{wch}$.
\begin{equation}
\hat{H}_S=\hbar(\omega_w|2\rangle\langle 2|+\omega_h|3\rangle\langle 3|)\,,
\end{equation}
and $\hat{S}_w=|1\rangle\langle 2|$, $\hat{S}_c=|2\rangle\langle 3|$, $\hat{S}_h=|1\rangle\langle 3|$.

\item[(iv)] $\mathcal{C}^{3+2m_h}$.
\begin{equation}
\hat{H}_S=\sum_{n=1}^{m_h+1} n\hbar\omega_h\,|2n+1\rangle\langle 2n+1|+ \hbar[ (n-1)\omega_h+\omega_c] \, |2n\rangle\langle 2n| \,,
\end{equation}
and $\hat{S}_c=|1\rangle\langle 2|$, $\hat{S}_w=|3+2m_h-1\rangle\langle 3+2m_h|$, $\hat{S}_h=\sum_{n=1}^{2m_h+1}|n\rangle\langle n+2|$.

\item[(v)] $\mathcal{C}^{3+3n_+}$.
\begin{equation}
\hat{H}_S=\sum_{n_h=0}^{n_++1}\sum_{n_c=0}^{\,n_+-n_h+1} \hbar[n_h\omega_h+n_c\omega_c] \,|n_h,n_c\rangle\langle n_h,n_c| \,,
\end{equation}
and $\hat{S}_c=\sum_{n_c=0}^{n_+} |0,n_c\rangle\langle 0,n_c+1|$, $\hat{S}_w=\sum_{n_h=0}^{n_+} |n_h,n_+-n_h+1\rangle\langle n_h+1,n_+-n_h|$, $\hat{S}_h=\sum_{n_h=0}^{n_+} |n_h,0\rangle\langle n_h+1,0|$.

\item[(vi)] $\mathcal{G}_3^B$.  
\begin{equation}
\hat{H}_S=\sum_{n=1}^{(N-1)/2} n\hbar\omega_h\,|2n+1\rangle\langle 2n+1|+ \hbar[(n-1)\omega_h+\omega_c] \, |2n\rangle\langle 2n| \,,
\end{equation}
and $\hat{S}_c=\sum_{n=1}^{(N-1)/2}|2n-1\rangle\langle 2n|$, $\hat{S}_w=\sum_{n=1}^{(N-1)/2} |2n\rangle\langle 2n+1|$,
$\hat{S}_h=\sum_{n=1}^{(N-1)/2} |2n-1\rangle\langle 2n+1|$.

\item[(vii)] $\mathcal{G}_{4}^B$.
\begin{equation}
\hat{H}_S=\sum_{n_h=0}^{\sqrt{N}-1}\sum_{n_c=0}^{\sqrt{N}-1} [n_h\omega_h+n_c\omega_c] \,|n_h,n_c\rangle\langle n_h,n_c| \,,
\end{equation}
and 
\begin{eqnarray}
\hat{S}_c &=& \sum_{n_h=0}^{\sqrt{N}-1}\sum_{n_c=0}^{\sqrt{N}-2} 
f(n_c) |n_h,n_c\rangle\langle n_h,n_c+1| \,, \nonumber \\
\hat{S}_w &=& \sum_{n_h=1}^{\sqrt{N}-1}\sum_{n_c=0}^{\sqrt{N}-2} 
g(n_h,n_c)|n_h-1,n_c+1\rangle\langle n_h,n_c| \,, \nonumber \\\
\hat{S}_h &=& \sum_{n_h=0}^{\sqrt{N}-2}\sum_{n_c=0}^{\sqrt{N}-1} 
f(n_h) |n_h,n_c\rangle\langle n_h+1,n_c| \,, 
\end{eqnarray}
with $f,g=1$. The Hamiltonian and coupling operators for $\mathcal{G}_{HO}^B$ are recuperated for an infinity number of states $N$, $f(n_\alpha)=\sqrt{n_\alpha+1}$, and $g(n_h,n_c)=\sqrt{n_h(n_c+1)}$.

\end{itemize}

\section{Hill theory and the steady state heat currents}\label{sec:appendixdecomposition}

\begin{figure}[h]
\centering
\includegraphics[width=\linewidth]{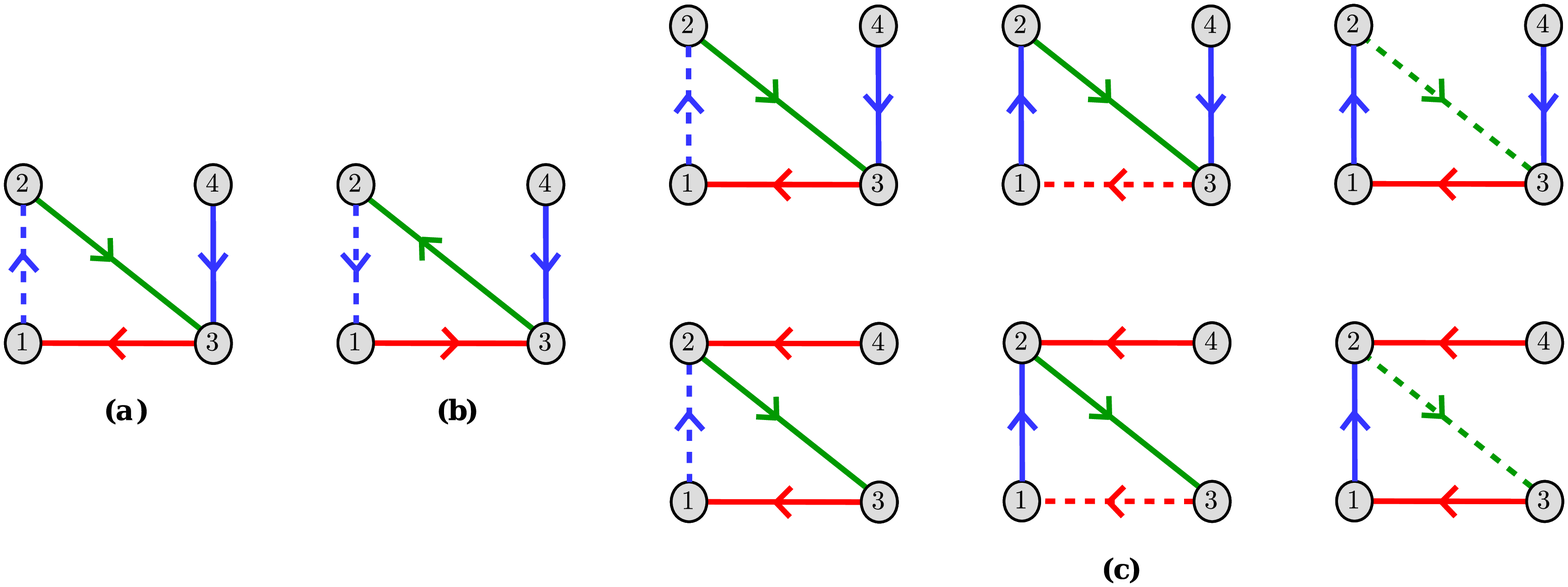}
\caption{Each term between brackets in equation (\ref{eq:entropy2}) is related to two subgraphs, as for example (a) $\vec{\mathcal{T}}^1_{1}+\vec{x}_1$ and (b) $\vec{\mathcal{T}}^1_{2}-\vec{x}_1$ of $\mathcal{G}_4$. When removing the cycles, the same forest, in this case the directed edge from vertex 4 to 3, remains. (c) Six different oriented maximal trees (solid lines). When adding the appropriate chord (dashed lines) the same cycle $\{1,2,3,1\}$ is obtained but with two different forests.
}
\label{fig:figC1}
\end{figure}
We apply Hill theory \cite{Hill1966} to obtain (\ref{eq:cycleheatcurrent}). The starting point is the steady state probability of finding the system in the state $i$ \cite{Schnakenberg1976,Hill1966}
\begin{equation}\label{eq:sspopulations}
p_i^s = D(\mathcal{G})^{-1}\sum_{\mu=1}^{N_T}\,\mathcal{A}(\vec{\mathcal{T}}_i^{\mu})\,,
\end{equation}
with $D$ given by (\ref{eq:determinant}). Introducing the steady state fluxes along a directed edge
\begin{equation}\label{eq:edgeflux}
J(\vec{x}_e) = W_{j_ei_e}^{\alpha_e} p_{i_e}^s
-W_{i_ej_e}^{\alpha_e} p_{j_e}^s\,,
\end{equation}
and the corresponding affinities
\begin{equation}
X(\vec{x}_e) = k_B\,\ln\left(\frac{W_{j_ei_e}^{\alpha_e}p_{i_e}^s}{W_{i_ej_e}^{\alpha_e} p_{j_e}^s}\right)\,,
\end{equation}
the total steady state entropy production is given by \cite{Broek2015,Schnakenberg1976}
\begin{equation}\label{eq:entropy1}
\dot{S} = \sum_{e=1}^{U} J(\vec{x}_e)X(\vec{x}_e)\,,
\end{equation}
where the orientation of each edge is arbitrary. Introducing the  populations in the product between fluxes and affinities
\begin{equation}
J(\vec{x}_e)X(\vec{x_e}) = 
D(\mathcal{G})^{-1}\sum_{\mu \in M_e}[W_{j_ei_e}^{\alpha_e}\mathcal{A}(\vec{\mathcal{T}}_{i_e}^\mu)-
W_{i_ej_e}^{\alpha_e}\mathcal{A}(\vec{\mathcal{T}}_{j_e}^\mu)]X(\vec{x}_e)\,,
\end{equation}
where $\sum_{\mu \in M_e}$ denotes the summation only over the maximal trees for which $x_e$ is a chord, since otherwise the term between brackets is zero. The product $W_{j_ei_e}^{\alpha_e}\mathcal{A}(\vec{\mathcal{T}}_{i_e}^\mu)$ is no more than the algebraic value $\mathcal{A}$ of the oriented subgraph $\vec{\mathcal{T}}_{i_e}^\mu+\vec{x}_e$, composed of the maximal tree $\vec{\mathcal{T}}_{i_e}^\mu$ and its chord $\vec{x}_e$. Then the entropy production (\ref{eq:entropy1}) can be written as
\begin{equation}\label{eq:entropy2}
\dot{S} = D(\mathcal{G})^{-1} \sum_{e=1}^{U} \sum_{\mu \in M_e}
[\mathcal{A}(\vec{\mathcal{T}}_{i_e}^\mu+\vec{x}_e) -\mathcal{A}(\vec{\mathcal{T}}_{j_e}^\mu-\vec{x}_e)]
X(\vec{x}_e)\,.
\end{equation}

Each term between brackets is only related to a circuit oriented in the two possible directions, $\vec{\mathcal{C}}_\nu$ and $-\vec{\mathcal{C}}_\nu$, associated with $\vec{\mathcal{T}}_{i_e}^\mu+\vec{x}_e$ and $\vec{\mathcal{T}}_{j_e}^\mu-\vec{x}_e$ respectively. When removing these two cycles from the corresponding subgraphs, the same forest $\vec{\mathcal{F}}_\nu^\beta$ remains, see for example figure \ref{fig:figC1}(a) and (b). Using this result and the properties of $\mathcal{A}$, each term in (\ref{eq:entropy2}) can be written as $\mathcal{A}(\vec{\mathcal{F}}_\nu^\beta)[\mathcal{A}(\vec{\mathcal{C}}_\nu)\,-\,\mathcal{A}(-\vec{\mathcal{C}}_\nu)]X(\vec{x}_e)$. The number of such terms with the same forest $\vec{\mathcal{F}}_\nu^\beta$ equals the number of edges of the circuit ${C}_\nu$, as shown in figure \ref{fig:figC1}(c). Next we introduce the cycle affinity (\ref{eq:cycleaffinity}), $X(\vec{\mathcal{C}}_\nu) = \sum_{e\in \nu} X(\vec{x}_e)$ with $\sum_{e\in \nu}$ the summation over all edges of $\vec{C}_\nu$, to obtain
\begin{equation}\label{eq:entropy3}
\dot{S} = D(\mathcal{G})^{-1} \sum_{\nu=1}^{N_C} 
\sum_{\beta \in \nu}\mathcal{A}(\vec{\mathcal{F}}_\nu^\beta)
[\mathcal{A}(\vec{\mathcal{C}}_\nu)-\mathcal{A}(-\vec{\mathcal{C}}_\nu)]
X(\vec{\mathcal{C}}_\nu)\,,
\end{equation}
where $\sum_{\beta \in \nu}$ denotes the summation over all the different forests associated with $\mathcal{C}_\nu$. This expression can be further simplified applying the matrix-tree theorem \cite{Moon1994}, $\sum_{\beta \in \nu}\mathcal{A}(\vec{\mathcal{F}}_\nu^\beta) = \det(-\mathbf{W}|\mathcal{C}_\nu)$. The flux associated with each cycle is
\begin{equation}\label{eq:cycleflux}
I(\vec{\mathcal{C}}_\nu) = D(\mathcal{G})^{-1}\det(-\mathbf{W}|\mathcal{C}_\nu)
[\mathcal{A}(\vec{\mathcal{C}}_\nu) - \mathcal{A}(-\vec{\mathcal{C}}_\nu)]\,.
\end{equation}
Considering that the cycle affinity and flux are odd functions, $X(-\vec{\mathcal{C}}_\nu)=-X(\vec{\mathcal{C}}_\nu)$ and $I(-\vec{\mathcal{C}}_\nu)=-I(\vec{\mathcal{C}}_\nu)$, we can define without any ambiguity the entropy production in the steady state corresponding to each circuit as
\begin{equation}\label{eq:cycleentropy}
\dot{s}(\mathcal{C}_\nu) = I(\vec{\mathcal{C}}_\nu)X(\vec{\mathcal{C}}_\nu)\ge 0\,,
\end{equation}
where the last inequality results from $D(\mathcal{G})^{-1}>0$, $\det(-\mathbf{W}|\mathcal{C}_\nu)>0$ and $[\mathcal{A}(\vec{\mathcal{C}}_\nu)-\mathcal{A}(-\vec{\mathcal{C}}_\nu)]\ln[\mathcal{A}(\vec{\mathcal{C}}_\nu)/\mathcal{A}(-\vec{\mathcal{C}}_\nu)]\ge 0$. Since the only contribution to the steady state entropy production is due to finite-rate heat transfer effects, we use (\ref{eq:cycleentropy}) to identify the circuit heat currents (\ref{eq:cycleheatcurrent}).

\section{Circuit decomposition of the four-state model}\label{sec:appendixfourstates}

Here we work out the circuit decomposition of $\mathcal{G}_{4}$. Now we only assume $E_1<E_2<E_3<E_4$, the consistency relation $E_{23}=E_{24}-E_{43}=E_{13}-E_{12}$ and the condition (\ref{eq:nondiag}). The transition matrix for the four-state model is given by
\begin{equation}
\mathbf{W}\,=\,\left(\begin{array}{cccc}
W_{11} & W_{12}^c & W_{13}^h & 0 \\
W_{21}^c & W_{22} & W_{23}^w & W_{24}^h \\
W_{31}^h & W_{32}^w & W_{33} & W_{34}^c \\
0 & W_{42}^h & W_{43}^c & W_{44}
\end{array}\right)\,,
\end{equation}
with diagonal elements $W_{11}=-W_{21}^c-W_{31}^h$, $W_{22}=-W_{12}^c-W_{32}^w-W_{42}^h$, $W_{33}=-W_{13}^h-W_{23}^w-W_{43}^c$ and $W_{44}=-W_{24}^h-W_{34}^c$. 

We denote by $\vec{\mathcal{C}}_1$ the cycle $\{1,2,3,1\}$, see figure \ref{fig:fig1}(d), for which using (\ref{eq:algebraic1}) we obtain $\mathcal{A}^c(\vec{\mathcal{C}}_1)=W_{21}^c$, $\mathcal{A}^w(\vec{\mathcal{C}}_1)=W_{32}^w$, and $\mathcal{A}^h(\vec{\mathcal{C}}_1)=W_{13}^h$. Then $\mathcal{A}(\vec{\mathcal{C}}_1)=W_{21}^c W_{32}^w W_{13}^h$ and $\mathcal{A}(-\vec{\mathcal{C}}_1)=W_{12}^c W_{23}^w W_{31}^h$. The cycle affinities associated with each bath (\ref{eq:cycleaffinitybath}) are $X^c(\vec{\mathcal{C}}_1)=E_{21}/T_c$ , $X^w(\vec{\mathcal{C}}_1)=E_{32}/T_w$, and $X^h(\vec{\mathcal{C}}_1)=E_{13}/T_h$, where $E_{ij}=E_j-E_i$. The contribution of the forests is $\det(-\mathbf{W}|C_1)=W_{24}^h+W_{34}^c$, from which the cycle flux is given by
\begin{equation}
I(\vec{\mathcal{C}}_1)=D(\mathcal{G}_4)^{-1}\,(W_{24}^h+W_{34}^c)
(W_{21}^c W_{32}^w W_{13}^h-W_{12}^c W_{23}^w W_{31}^h)\,,
\end{equation}
where $D(\mathcal{G}_4)$ is determined by using  (\ref{eq:determinant}). Then the circuit heat currents are $\dot{q}_c(\mathcal{C}_1)=E_{12}I(\vec{\mathcal{C}}_1)$,
$\dot{q}_w(\mathcal{C}_1)=E_{23}I(\vec{\mathcal{C}}_1)$ and $\dot{q}_h(\mathcal{C}_1)=E_{31}I(\vec{\mathcal{C}}_1)$. The consistency of the circuit currents with the first law $\dot{q}_c(\mathcal{C}_1)+\dot{q}_w(\mathcal{C}_1)+\dot{q}_h(\mathcal{C}_1)=0$ follows from $E_{12}+E_{23}+E_{31}=0$. The cycle affinity  (\ref{eq:cycleaffinity}) is $X(\vec{\mathcal{C}}_1)=E_{21}/T_c+E_{32}/T_w+E_{13}/T_h$ from which the circuit entropy production can be determined with (\ref{eq:cycleentropy}). A similar procedure can be used in order to obtain the quantities associated with the circuit $\mathcal{C}_2$.

For the circuit $\mathcal{C}_3$ we denote by $\vec{\mathcal{C}}_3$ the cycle $\{1,2,4,3,1\}$. Now $\mathcal{A}^c(\vec{\mathcal{C}}_3)=W_{21}^c W_{34}^c$, $\mathcal{A}^w(\vec{\mathcal{C}}_3)=1$ (there is not any edge associated with the work bath) and $\mathcal{A}^h(\vec{\mathcal{C}}_3)=W_{42}^h W_{13}^h$, $\mathcal{A}(\vec{\mathcal{C}_3})=W_{21}^c W_{34}^c W_{42}^h W_{13}^h$ and $\mathcal{A}(-\vec{\mathcal{C}_3})=W_{12}^c W_{43}^c W_{24}^h W_{31}^h$. The cycle affinities associated with each bath are $X^c(\vec{\mathcal{C}}_3)=(E_{34}-E_{12})/T_c$, $X^w(\vec{\mathcal{C}}_3)=0$ and $X^h(\vec{\mathcal{C}}_3)=(E_{13}-E_{24})/T_h$. Notice that $(E_{13}-E_{24})=-(E_{34}-E_{12})$. When the transition energies are equal, $E_{34}=E_{12}$ and $E_{24}=E_{13}$, all the affinities are zero. The circuit $\mathcal{C}_3$ involves all the graph vertices and therefore there is not any forest associated with it. Then $(-\mathbf{W}|C_3)$ is an empty matrix and $\det(-\mathbf{W}|C_3)=1$. The cycle flux is given by
\begin{equation}
I(\vec{\mathcal{C}}_3)=D(\mathcal{G}_4)^{-1}\,
(W_{21}^c W_{34}^c W_{42}^h W_{13}^h - W_{12}^c W_{43}^c W_{24}^h W_{31}^h)\,,
\end{equation}
and the circuit heat currents by $\dot{q}_c(\mathcal{C}_3)=(E_{43}-E_{21})I(\vec{\mathcal{C}}_3)$,
$\dot{q}_w(\mathcal{C}_3)=0$ and $\dot{q}_h(\mathcal{C}_3)=(E_{31}-E_{42})I(\vec{\mathcal{C}}_3)$.


\section{Other decompositions of the entropy production}\label{sec:appendixSchnakenberg}

There are several possible decompositions of the steady state entropy production in terms of circuits. Schnakenberg \cite{Schnakenberg1976} designed a method based on the identification of a set of $U-N+1$ fundamental circuits. The circuits are determined by choosing an arbitrary maximal tree and adding each one of its chords. Taking a particular orientation for the circuits, a set of fundamental cycles is found. The total steady state entropy production is then $\dot{S}=\sum_{\nu=1}^{U-N+1}J(\vec{x}_\nu)X(\vec{\mathcal{C}}_\nu)$, where $x_\nu$ is the chord giving the circuit $\mathcal{C}_\nu$ and $J(\vec{x}_\nu)$ the corresponding flux. The previous decomposition is simple and specially relevant when $U-N$ is small. However, it is not unique, since it depends on the choice of the maximal tree, and some terms in the sum may be not positive definite, which discards a possible consistent thermodynamic interpretation of each circuit contribution. Besides the evaluation of $J(\vec{x}_\nu)$ requires the calculation of the steady state populations.

As an example we apply Schnakenberg method to the four-state model. The procedure requires an arbitrary set of fundamental circuits of the graph $\mathcal{G}_4$. We choose the maximal tree shown in figure \ref{fig:fig2}(a), which has two chords, $\{1,2\}$ and $\{2,4\}$. By adding the chord $\{1,2\}$ the circuit $\mathcal{C}_1$ is obtained. Choosing an arbitrary orientation, for example $\vec{\mathcal{C}}_1$ as in figure \ref{fig:fig2}(d), the directed chord $\vec{x}_1$ goes from state 1 to state 2. In this decomposition the flux associated with each cycle is taken as the corresponding to the directed chord (\ref{eq:edgeflux}), $J(\vec{x}_1) =\,W_{21}^c p^s_1 - W_{12}^c p^s_2$. The cycle affinity is defined by (\ref{eq:cycleaffinity}) and was calculated in \ref{sec:appendixfourstates}, $X(\vec{\mathcal{C}}_1)=E_{21}/T_c+E_{32}/T_w+E_{13}/T_h$. When adding the chord $\{2,4\}$ we obtain the circuit $\mathcal{C}_2$. Choosing the orientation $\{2,3,4,2\}$, the directed chord $\vec{x}_2$ goes from state 4 to state 2. The flux is $J(\vec{x}_2)\,=\,W_{24}^h p^s_4 - W_{42}^h p^s_2$ and the affinity $X(\vec{\mathcal{C}}_2)=E_{43}/T_c+E_{32}/T_w+E_{24}/T_h$. Then the entropy production is
\begin{equation}\label{eq:SEP}
\dot{S}\,=\,J(\vec{x}_1)X(\vec{\mathcal{C}}_1)+J(\vec{x}_2)X(\vec{\mathcal{C}}_2)\,.
\end{equation}
The cycles $\{\vec{\mathcal{C}}_1,\vec{\mathcal{C}}_2\}$ are the elements of one of the possible fundamental sets of $\mathcal{G}_4$.  Notice that for our choice the circuit $\mathcal{C}_3$ is not involved. 

A related decomposition is obtained by the algorithm of Kalpazidou \cite{MacQueen1981,Kalpazidou2007}. For systems showing dynamical reversibility the algorithm leads to a Schnakenberg decomposition with a clever choice of the fundamental set of cycles, such that all the terms in the sum are positive. Therefore a positive entropy production can be assigned to each cycle, which is required in many applications \cite{Altaner2012b,Knoch2015}. The algorithm is based on choosing an orientation for the graph such that all the fluxes (\ref{eq:edgeflux}) for the directed edges are positive. Next a cycle is identified and the entropy production $J_{min}(\vec{x}_\nu) X(\vec{\mathcal{C}}_\nu)>0$ is assigned to it, where $J_{min}(\vec{x}_\nu)$ is the smallest flux associated with an edge of $\vec{\mathcal{C}}_\nu$. Then $J_{min}(\vec{x}_\nu)$ is subtracted to each flux in the cycle to obtain a new flux field and the process is repeated for new cycles until a fundamental set is completed \cite{Altaner2012b,Knoch2015}. For example, let us assume parameter values for which the two triangles of the four-state model work as refrigerators. Then the fluxes along $\vec{x}_1$, $\vec{x}_2$, $\vec{x}_3$ (from 3 to 4), $\vec{x}_4$ (from 3 to 1) and $\vec{x}_5$ (from 2 to 3) are positive. With this orientation only the cycles $\vec{\mathcal{C}}_1$ and $\vec{\mathcal{C}}_2$ appear in the directed graph and the entropy production can be written as (\ref{eq:SEP}), where the two terms are guaranteed to be positive. If we modify the system parameters such that the circuit $\mathcal{C}_2$ works as a heat transformer but the overall device remains working as a refrigerator, the total entropy production can still be determined using (\ref{eq:SEP}), but the positivity of each term is not guaranteed. Now the fluxes are positive along the edges $\vec{x}_1$, $-\vec{x}_2$, $-\vec{x}_3$, $\vec{x}_4$  and $\vec{x}_5$. Only the cycles $\vec{\mathcal{C}}_1$ and $\vec{\mathcal{C}}_3$ remain with this graph orientation and the algorithm of Kalpazidou leads to
\begin{equation}
\dot{S}\,=\,J(\vec{x}_5)X(\vec{\mathcal{C}}_1)+J(-\vec{x}_2)X(\vec{\mathcal{C}}_3)\,.
\end{equation}
However, in this expression the contribution of each mechanism, refrigerator ($\mathcal{C}_1$), heat transformer ($\mathcal{C}_2$) and heat leak ($\mathcal{C}_3$) could not be isolated.

\section{$\mathcal{R}(\vec{\mathcal{C}}^{\,N})$ in the high and low temperatures limits}\label{sec:appendixR}

In the high temperature limit, $y_\alpha\equiv\exp[-E_\alpha/(k_BT_\alpha)]\approx 1$, the transition rates satisfy $W_{-\alpha}\approx W_{\alpha}$, leading to vanishing affinities and heat currents. Now $\mathcal{A}(\vec{\mathcal{T}}_i^j)$ is in a good approximation independent of the orientation, what considerably facilitates the calculations to obtain 
\begin{equation}\label{eq:hightemperatures}
\mathcal{R}(\vec{\mathcal{C}}^{\,N}) \approx
\left[N\left(\frac{r_c}{W_c}+\frac{r_h}{W_h}+\frac{r_w}{W_w}\right)\right]^{-1} \,,
\end{equation}
with $r_\alpha=n+m_\alpha$ and $r_c+r_w+r_h=N$. In this limit $\mathcal{R}(\vec{\mathcal{C}}^{\,N})$ decreases quadratically ($z=2$) with $N$, except when one or two of the terms $r_\alpha/W_\alpha$ are much larger than the others and $r_\alpha$ remains constant when increasing $N$, which can only happens adding two-edge sets. In this limit $\mathcal{R}(\vec{\mathcal{C}}^{\,N})$ decreases as $N^{-1}$ for small enough values of $N$.

In the low temperature limit, $y_\alpha\ll 1$ and $W_{-\alpha}\ll W_{\alpha}$. Again this limit implies vanishing heat currents. The cycle algebraic value is proportional to the small factors $y_\alpha$, $\mathcal{A}(\vec{\mathcal{C}}^{\,N})\propto \prod_\alpha y_\alpha^{u_\alpha+u_\alpha'}$, where $u_\alpha$  and $u_\alpha'$ are the number of $W_{-\alpha}$ transitions before and after the highest-energy state respectively. Besides, the largest contributions to $D$ comes from two terms that include the lowest number of rates $W_{-\alpha}$, $\mathcal{A}(\vec{\mathcal{T}}_1^{h-1})$ and $\mathcal{A}(\vec{\mathcal{T}}_1^{h})$, being $i=1$ the ground state and $j=h$ the highest-energy state. Both terms are proportional to $\prod_\alpha y_\alpha^{f_\alpha+u_\alpha'}$, where $f_\alpha$ is the number of $W_{\alpha}$ transitions before the highest-energy state in $\mathcal{A}(\vec{\mathcal{C}}^{\,N})$. Necessarily $u_\alpha-f_\alpha$ is positive, increases with $N$ and then $\mathcal{R}(\vec{\mathcal{C}}^{\,N})$ decreases exponentially when adding new states to the circuit. For example, $\mathcal{R}(\vec{\mathcal{C}}^{\,N})\propto \prod_\alpha \exp[-u_\alpha E_\alpha/(k_B T_\alpha)]$ when $f_\alpha=0$. Examples of these behaviors are given in figures \ref{fig:fig4}(b) and (e).

\section{Heat currents and spectral radius of $\mathcal{G}_3^{B}$}\label{sec:appendixG3B}

\begin{figure}[h]
\begin{minipage}{\linewidth}\centering
\includegraphics[width=0.6\linewidth]{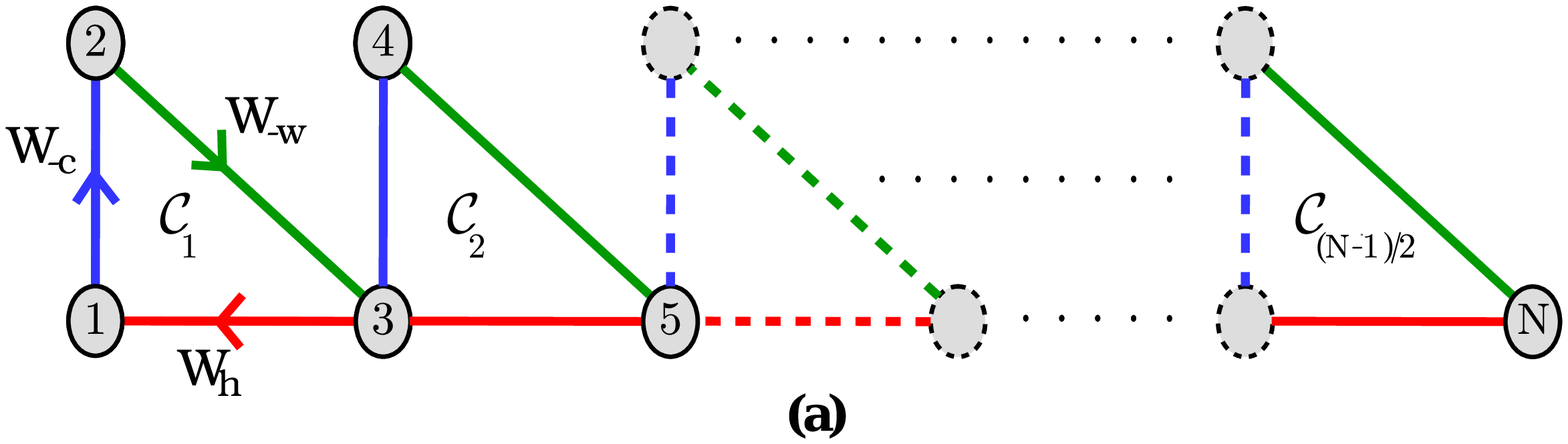}
\end{minipage}

\vspace*{0.5cm}

\begin{minipage}{\linewidth}\centering
\includegraphics[width=0.45\linewidth]{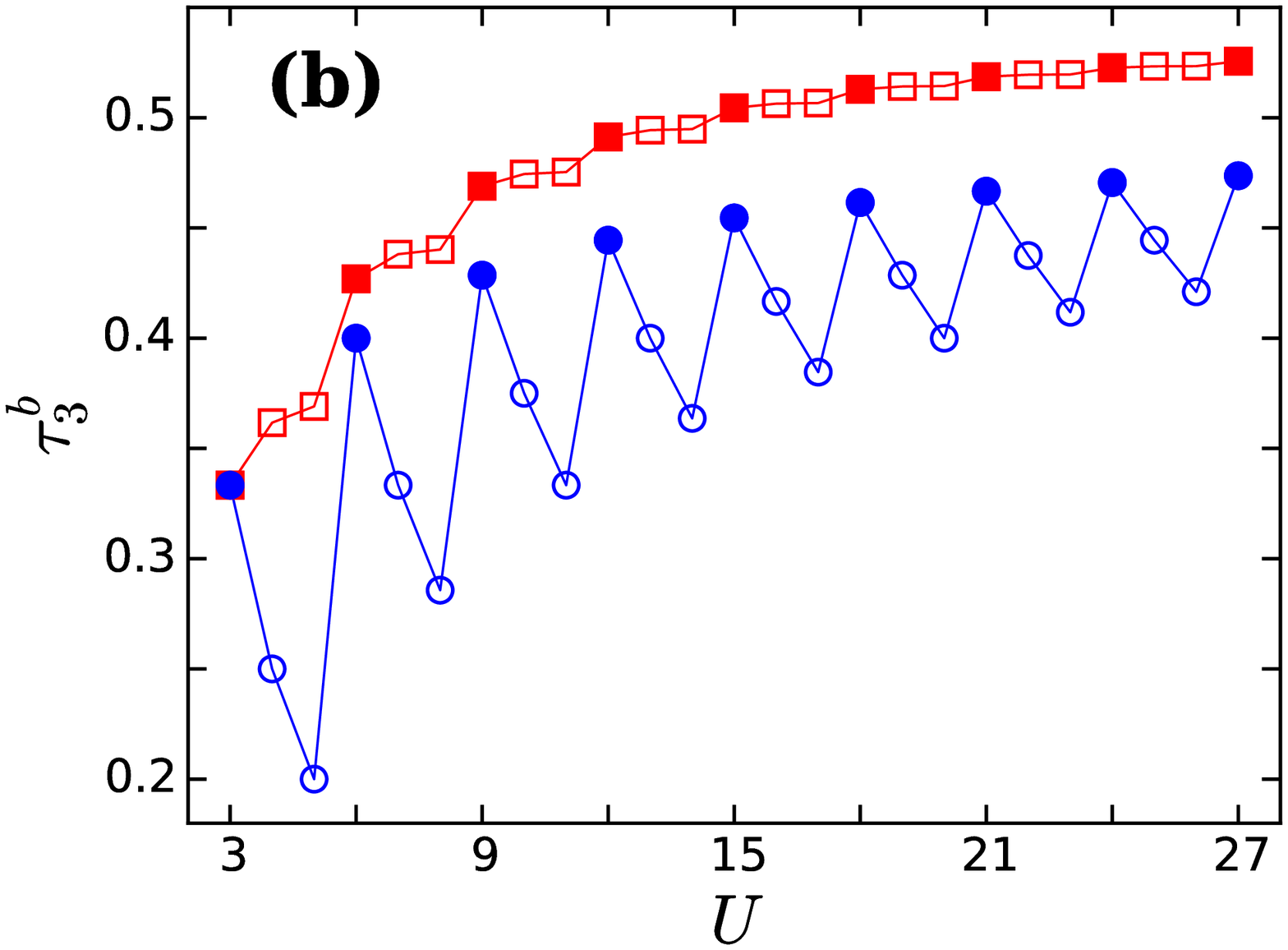}
\includegraphics[width=0.45\linewidth]{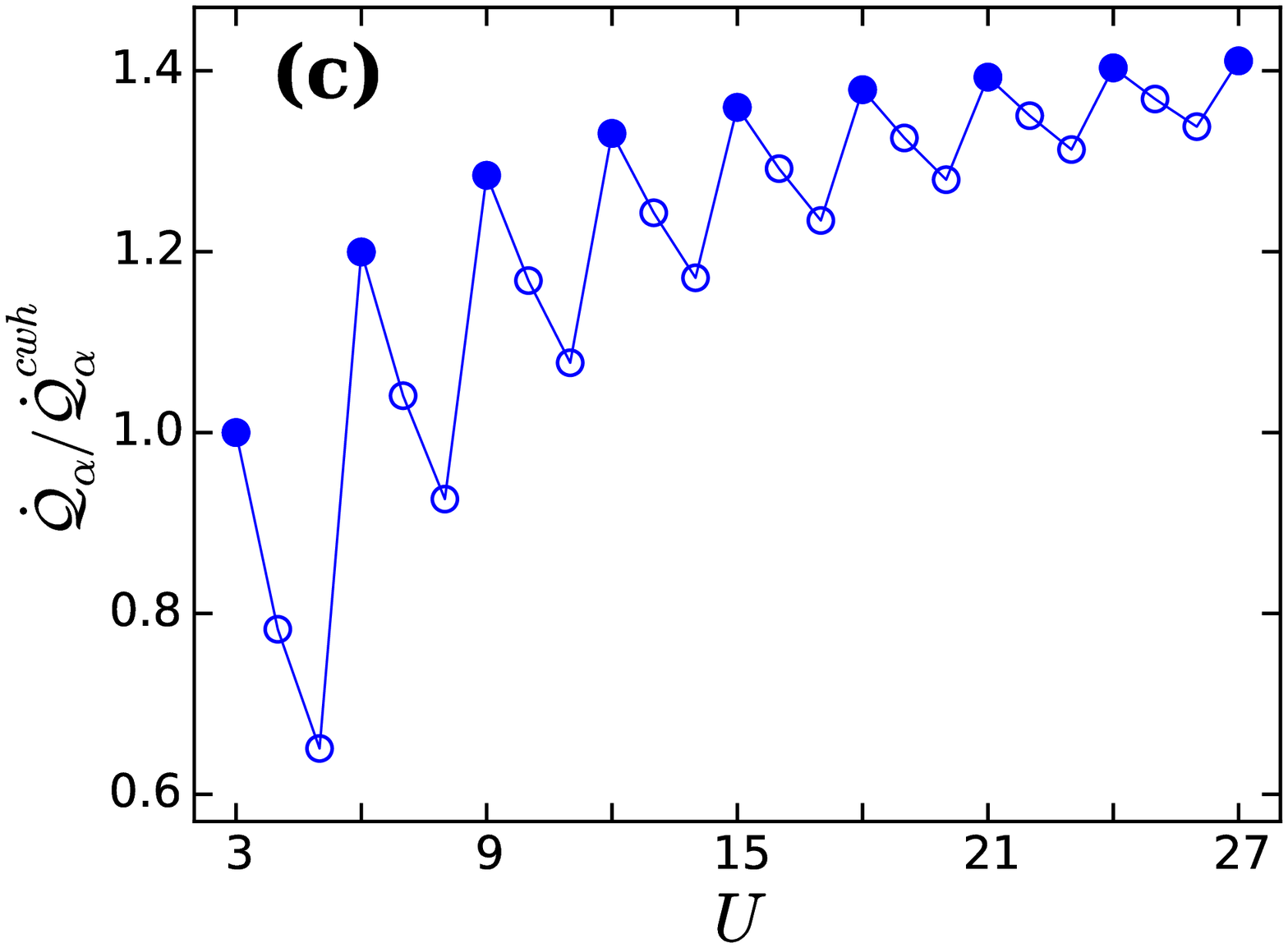}
\end{minipage}
\caption{(a) The graph $\mathcal{G}_3^B$. (b) The parameter $\tau_3^b$ (circles), the spectral radius $\rho(\mathbf{A})/6$ (squares) and (c) the physical heat currents (normalized to $\dot{\mathcal{Q}}_\alpha^{cwh}$)  as functions of the number of edges $U$. The parameters are the same as in figure \ref{fig:fig5} with $t=1$. Solid symbols are used when a new triangle is completed. The lines are merely eye guides}\label{fig:figG1}
\end{figure}

The simple topological structure of $\mathcal{G}_3^B$, see figure \ref{fig:figG1}(a), allows for the direct identification of all the $N_T=3^{N_C}$ maximal trees. Then
\begin{equation}
\det(-\mathbf{W}|\mathcal{C}_\nu)D(\mathcal{C}_3)=
\sum_{\mu=1}^{N_T}\sum_{i=2\nu-1}^{2\nu+1}\mathcal{A}(\vec{\mathcal{T}}_i^\mu)\,.
\end{equation}
Using this result the physical heat currents are given by
\begin{equation}
\dot{\mathcal{Q}}_\alpha=\left[
1+\frac{\sum_{\mu=1}^{N_T}\sum_{i=1}^{N_C-1}\mathcal{A}(\vec{\mathcal{T}}_{2i+1}^\mu)}
{\sum_{\mu=1}^{N_T}\sum_{i=1}^{N}\mathcal{A}(\vec{\mathcal{T}}_i^\mu)}\right]\dot{\mathcal{Q}}_\alpha^{cwh}
\equiv K \dot{\mathcal{Q}}_\alpha^{cwh}\,,
\end{equation}
where $1\leq K\leq 2$ and $K=3N_C/(2N_C+1)$ in the high temperature limit.

For this graph $\lambda(\mathcal{C}_\nu)=\frac{1}{3}$ and $\tau_3=\tau_3^b/3$. The parameter $\tau_3^b$ as a function of the number of edges is shown in figure \ref{fig:figG1}(b). The spectral radius $\rho(\mathbf{A})$, defined as the largest eigenvalue of the adjacency matrix of the unweighted graph \cite{Wilson1979}, is also shown. The spectral radius is a measure of the graph connectivity which increases monotonically with the number of edges. However, it does not reflect the decrease in the heat currents each time a pendant edge is added to the graph, see figure \ref{fig:figG1}(c). An increment in the total heat currents is only found  when a new triangle is completed, saturating to a constant value when the addition of new circuits does not improve significantly the graph connectivity. This behavior is well described by $\tau_3^b$.


\end{document}